\documentclass{emulateapj-rtx4}

\usepackage[normalem]{ulem}
\usepackage{color}

\usepackage{tikz}
\usetikzlibrary{shapes,arrows}

\newcommand{\be}{\begin{eqnarray}}
\newcommand{\ee}{\end{eqnarray}}

\usepackage{amsmath}

\begin{document}

\title{ Formation of Extremely Low-mass White Dwarfs Binaries  }

\author{M. Sun\altaffilmark{1}, P. Arras\altaffilmark{1}} 
 \altaffiltext{1}{Department of Astronomy, University of Virginia,
P.O. Box 400325, Charlottesville, VA 22904, USA}
\email{ msun@virginia.edu, arras@virginia.edu}

\begin{abstract}

Motivated by the discovery of a handful of pulsating, extremely low-mass white dwarfs (ELM WDs, mass $M \la 0.18\, M_\odot$) which likely have WD companions, this paper discusses binary formation models for these systems. ELM WDs are formed using angular momentum losses by magnetic braking. Evolutionary models are constructed using the Modules for Experiments in Stellar Astrophysics (MESA), with ELM WD progenitors in the range $1.0 \la M_{\rm d}/M_{\odot} \la 1.5$ and WD companions in the range $0.4 \la M_{\rm a}/M_{\odot} \la 0.9$. A prescription to reduce magnetic braking for thin surface convection zones is included. Upon the thinning of the evolved donor's envelope, the donor star shrinks out of contact and mass transfer (MT) ceases, revealing the ELM WD. Systems with small masses have previously been suggested as possible AM CVN's. Systems with large masses, up to the limit $M \simeq 0.18\, M_\odot$ at which shell flashes occur on the WD cooling track, tend to expand out to orbital periods $P_{\rm orb} \ga 15\, {\rm hr}$. In between this range, ELM WDs may become pulsators both as pre-WDs and on the WD cooling track. Brickhill's criterion for convective mode driving is used to estimate the location of the blue edge of the g-mode instability strip. In the appendix, we show that the formation of an ELM WD by unstable MT or a common envelope event is unlikely. Stable Roche-lobe overflow with conservative MT produces only $M \ga 0.2\, M_\odot$.

\end{abstract}



\section{Introduction}
\label{sec:intro}

Extremely low-mass white dwarfs (ELM WDs) are here defined as helium-core WDs with masses $M \la 0.18\ M_\odot$, sufficiently low that no hydrogen shell flashes occur during the WD cooling stage. More massive WDs have shell flashes, which quickly decrease the mass of the hydrogen-rich envelope. The thicker envelopes of ELM WDs allow for significantly higher stable hydrogen burning rates, keeping these lower mass stars more luminous than their slightly more massive counterparts \citep{1999A&A...350...89D}.

Helium core WDs can in principle be formed through single star evolution, for sufficiently small mass that helium core ignition is avoided. Large helium core WDs of mass $M \la 0.45 \, M_\odot$ may be produced in less than 13.7 Gyr \citep{1996ApJ...466..359D} if enhanced mass-loss rates are assumed on the red giant branch (RGB). However, far larger mass-loss rates would be needed to strip more than $0.8\, M_\odot$ on the RGB to uncover a helium core mass $M_{\rm c}<0.2\, M_\odot$ starting from a $M_i \simeq 1\, M_\odot$ zero-age main sequence (ZAMS) star. Further, the main sequence (MS) evolution time for ZAMS masses $M_{\rm i} \la 0.2\, M_\odot$ is $t_{\rm ms} \ga 1000\, \rm Gyr$, and hence, in practice, ELM WDs can only be produced through binary evolution, either by stable Roche-lobe overflow (RLOF) or unstable mass transfer (MT) and a common envelope (CE) inspiral.

ELM WDs have been observed to pulsate with g and p-mode oscillations \citep{2012ApJ...750L..28H,2013ApJ...765..102H,2013MNRAS.436.3573H}, opening up the possibility of probing the interiors of these exotic stars with seismology. One question is whether the two different formation channels can be distinguished through seismology. Since the structure of these objects is particularly simple, with a helium core and a thick, hydrogen-rich envelope, there are in principle fewer parameters required to characterize the star than for carbon-oxygen core DA and DB WD pulsators. One complication is that, for the effective temperature $T_{\rm eff}$ range of observed pulsators, there may be insufficient time to establish diffusive equilibrium throughout the star \citep{2014A&A...569A.106C}. This complicates the calculation of stellar models, because time-dependent diffusion must be included, but also provides an additional opportunity for changing composition profiles to affect the mode periods.

A number of ELM WDs were recently discovered by the ELM, SPY and WASP surveys \citep{2009A&A...505..441K, 2010ApJ...723.1072B, 2012ApJ...744..142B, 2013ApJ...769...66B, 2016ApJ...818..155B, 2011MNRAS.418.1156M, 2011ApJ...727....3K, 2012ApJ...751..141K, 2015ApJ...812..167G}. Follow-up observations of ELM survey candidates allowed \citet{2012ApJ...750L..28H} to discover the first pulsating ELM WD, SDSS J184037.78+642312.3. Subsequently, the 2nd and 3rd pulsating helium WDs, SDSS J111215.82+111745.0 (J1112) and SDSS J151826.68+065813.2, were discovered with seven oscillation frequencies, respectively. For J1112, two modes with shorter periods were suggested as possible p-mode oscillations \citep{2013ApJ...765..102H}. This is the only known WD with p-mode pulsations. There are currently seven known pulsating ELM WDs (see Table \ref{properties_seven_pulsators}). The $T_{\rm eff}$ range of the seven stars is $7,890 < T_{\rm eff}/{\rm K} < 9,560$, and the surface gravity range is $5.78 < {\rm log}_{10}\,(g/{\rm cm\ s^{-2}}) < 6.68$. The range of observed pulsation periods for the seven objects is from 1184 to 6235s \citep{2013MNRAS.436.3573H,2015ASPC..493..217B,2015MNRAS.446L..26K}. Pulsations are also observed in pre-ELM WDs WASP J1628+10B and WASP J0247-25B, which have not yet cooled off to the WD cooling track. An interesting point in regards to the driving of the observed modes by the $\kappa$ mechanism is that the driving by the helium partial ionization zone may indeed explain the observed pulsations \citep{2016A&A...588A..74C}, but it was necessary to turn off diffusion, otherwise helium would settle down below the driving region. Models which include element diffusion \citep{2016A&A...595A..35I} must include a source of mixing to keep the helium lofted up in the driving region.

The expected range of oscillation mode periods of helium WDs with mass $M \la 0.2M_{\odot}$ was examined by \citet{2010ApJ...718..441S}, who showed that the smaller WD mass and larger radius lead to mode periods as much as a factor of 2 longer than the carbon-oxygen core WD with $\log_{10}\,(g/{\rm cm\,s^{-2}}) \approx 8$. They also showed that g-mode pulsations may contain most of their energy in the helium core, so that the mode periods may be sensitive to $M_{\rm c} $. C{\'o}rsico and Althaus (2014; hereinafter CA) studied the two short period p-modes of J1112, finding that the model p-modes nearly match the observed short period modes for a low-mass WD with $M \simeq 0.16\ M_\odot$, but the implied surface gravity was then well below that inferred from spectra. Subsequently \citet{2015ApJ...809..148T} used three-dimensional (3-D) hydrodynamic simulations of WD atmospheres, and fitting these new model atmospheres implied significantly different ${\rm log}\,g$ for cool DA WDs, as much as 0.35 dex, closer than that required by the short p-mode periods. At present, the minimum mass ELM WD from the ELM survey is $M=0.14\ M_{\odot}$ \citep{2016ApJ...818..155B} with a $\log_{10}\,(g/{\rm cm\,s^{-2}}) = 5.5$. One question addressed in this paper is the minimum mass for the ELM WD from binary evolution.

There are seven observed pulsating WDs with mass lower than $0.2M_{\odot}$. Table~\ref{properties_seven_pulsators} gives their parameters from observations. The mass estimates of the ELM WDs and their companions are shown in columns 2 and 6. Except J1618+3854, all other ${\rm log}\,g$ is given by 3-D atmosphere simulations \citep{2015ApJ...809..148T,2015MNRAS.446L..26K}. Three of the systems have no radial velocity detection of a companion.
\begin{center}
\begin{table*}[]
\caption{Properties of the seven pulsating ELM WDs}
\begin{center}
\begin{tabular}{c c c c c c c c}

\hline\hline 
Object	&	$M$&	$\log_{10}\,g$&	$T_{\rm eff}$&	Mass Function&	$M_{\rm 2,min}$&	$P_{\rm orb}$ & Ref.\\
		&	($M_{\odot}$)&	(${\rm cm\,s^{-2}}$)&	(K)&	($M_{\odot}$)&	($M_{\odot}$)&	(hrs) & \\
\hline
J1840+6423&	0.177&	6.34 $\pm$ 0.05&	9120 $\pm$ 140&	0.399 $\pm$ 0.009&	0.65 $\pm$ 0.03&	4.5912 $\pm$ 0.001 & (1)(7)\\
J1112+1117&	0.169&	6.17 $\pm$ 0.06&	9240 $\pm$ 140&	0.028 $\pm$ 0.003&	0.14 $\pm$ 0.01&	4.1395 $\pm$ 0.0002 & (2)(7)\\
J1518+0658&	0.197&	6.68 $\pm$ 0.05&	9650 $\pm$ 140&	0.322 $\pm$ 0.005&	0.58 $\pm$ 0.03&	14.624 $\pm$ 0.001 & (2)(7)\\
J1614+1912&   	0.172&	6.32 $\pm$ 0.13&	8700 $\pm$ 170&	...&...&...& (3)(7)\\
J2228+3623&   	0.175&	5.78 $\pm$ 0.08&	7890 $\pm$ 120&	...&...&...& (3)(7)\\
J1618+3854&	0.179&	6.54 $\pm$ 0.14&	8965 $\pm$ 120	&	...&...&...& (4)\\		
J1738+0333&	0.172&	6.30 $\pm$ 0.10&	8910 $\pm$ 150&	0.0003455012&	1.47 $\pm$ 0.07&	8.51496   & (5)(6)\\
\hline
\label{properties_seven_pulsators}
\end{tabular}

(1)\cite{2012ApJ...750L..28H}; (2)\cite{2013ApJ...765..102H}; (3)\cite{2013MNRAS.436.3573H}; (4)\cite{2015ASPC..493..217B}; (5)\cite{2015ApJ...812..167G}; (6)\cite{2015MNRAS.446L..26K};  (7)\cite{2015ApJ...809..148T}.
\end{center}
\end{table*}
\end{center}

Section \ref{sec:elm_from_mb} discusses a promising formation channel, Roche-lobe overflow including orbital angular momentum losses due to magnetic braking.  Binary evolution and ELM formation results from this model are presented in Section \ref{sec:results}. Discussion and conclusions are given in Sections \ref{sec:discussion} and \ref{sec:conclusions}. Appendix \ref{sec:conservative} shows that the minimum WD mass produced through conservative mass transfer is larger than the ELM WD mass range. Appendix \ref{sec:ce} shows that formation of an ELM WD by CE evolution tends to produce very close binaries, which may merge in many cases.

\section{ ELM formation through magnetic braking}
\label{sec:elm_from_mb}

The Cataclysmic Variable (CV) model of ELM WD formation in this paper assumes that the progenitor of the ELM WD was the initially less massive star. The initially more massive star formed a WD companion. 

From the discussion in Appendix \ref{sec:conservative}, magnetic braking is key to form ELM WDs so that the envelope is stripped before the core can grow too large. This section starts with a brief summary of previous work on CV binaries with both unevolved ($M_{\rm c}=0$) and slightly evolved ($M_{\rm c} \la 0.05\, M_\odot$)
stars transferring mass to a WD. The ELM formation model presented here is the extension to CVs with higher core masses in the range $0.06 \la M_{\rm c}/M_{\odot} \la 0.16$. In this model, the ELM WD started as the donor star, and appeared as a $M<0.18\, M_\odot$ WD at the end of MT. The lower core mass end for the ELM WD comes from the requirement that MT ends before the AM CVN phase, so that the ELM WD may be observed as a pulsator. The upper core mass limit for the ELM WD is set by requiring that no shell flashes occur on the WD cooling track, allowing thick surface hydrogen layers and long-lived stable nuclear burning.

Canonical CV evolution of unevolved donor stars with masses $M_{\rm d} \la 1\, M_\odot$ uses magnetic braking laws calibrated by observations of the spin-down of single stars to understand binary evolution. Since the thermal time is shorter than the mass-loss timescale for these systems, the evolution is relatively insensitive to the initial donor mass, and the evolution of different donor masses converges to the same track at shorter $P_{\rm orb}$. The well-known CV period gap, the scarcity of accreting systems in the range $2<P_{\rm orb}/{\rm hr}<3$, is understood as the donor shrinking inside the Roche lobe when the magnetic braking torque decreases sharply. The physical origin of the angular momentum loss rate by magnetic braking $\dot{J}_{\rm mb}$ was initially thought to be the disappearance of the tachocline as the star became fully convective, although it was later realized that even late M stars may be able to generate large magnetic fields which can support a comparable level of coronal activity required to generate a magnetic wind \citep{1967ApJ...150..551K,1972ApJ...171..565S,1983A&A...124..267S,2008ApJ...676.1262B}. Regardless of the origin of the torque decrease, it is implemented in evolutionary codes by turning $\dot{J}_{\rm mb}$ off by hand when the donor star becomes fully convective. MT then resumes at $P_{\rm orb} \simeq 2\, {\rm hr}$ when gravitational wave torques shrink the orbit and bring the donor back into contact. The gradual lengthening of the thermal time as the hydrogen burning limit is approached changes the structure of the donor from that of a low-mass MS star to a brown dwarf responding adiabatically to mass-loss. As a result the donor star expands upon losing mass, and the orbital evolution switches from contraction to expansion. 

The evolution of CVs with slightly evolved donors $M_{\rm c} \la 0.05\, M_\odot$ has been discussed by \citet{2003MNRAS.340.1214P} and \citet{2006A&A...460..209V}. They showed that systems with evolved donors can form short-period AM CVn systems for small $M_{\rm c}$, and also dominate the CV population at long orbital periods $P_{\rm orb} \ga 5\, \rm hr$ for larger $0.03\la M_{\rm c}/M_{\odot} \la 0.05$. A bifurcation period at $16\la P_{\rm orb}/{\rm hr} \la 22$ separates the systems which move to shorter periods from those that expand. In the period range $1\leqslant P_{\rm orb}/{\rm hr} \leqslant 5\, \rm hr$ the CV population is dominated by unevolved stars. 

\citet{2003MNRAS.340.1214P} discussed that, as compared to unevolved donors, care must be taken in the magnetic braking torque when the donor's convective envelope becomes thin. The commonly used $\dot{J}_{\rm mb}$ formulae have been calibrated for stellar masses less than about $1\, M_\odot$, and do not take into account the reduced magnetic torque for sufficiently thin surface convection zones. The well known Kraft break \citep{1967ApJ...150..551K} in the rotation rates of single stars at mass about $1.3\, M_\odot$ divides the higher-mass, rapid rotators from the lower-mass slow rotators, indicating a dramatic reduction in $\dot{J}_{\rm mb}$ when the surface convection zone becomes small. For evolved donors, this reduction is key to the formation of ELM WDs. Due to the degenerate helium core, these stars always have radiative cores, and hence $\dot{J}_{\rm mb}$ would not undergo the same drastic reduction as for unevolved donors. However, MT gradually sheds the envelope until it becomes so thin that the shell burning strongly decreases, with an associated shrinking of the convective envelope. This tends to cause the evolved donor to fall out of contact. If, in addition, a prescription for reduced $\dot{J}_{\rm mb}$ at small convective envelope mass is included then a long non-accreting phase in which the donor star emerges as an ELM WD may result. Systems with small $M_{\rm c}$ which fall out of contact at small $P_{\rm orb}$ may be driven back into contact by gravitational wave losses, while those with larger $M_{\rm c}$ are sufficiently distant that they don't have time to come back into contact in a Hubble time.

The magnetic braking law chosen here is the same as used for unevolved donors, with a reduction in torque for small convective envelopes. The reasonableness of this prescription can be judged by the agreement of the model $P_{\rm orb}$, $\log_{10}\,g$ and $T_{\rm eff}$ with observations.

\subsection{ Description of the Simulations}

Binaries are evolved using the ``binary\_donor\_only" option in the Modules for Experiments in Stellar Astrophysics code (MESA, version 8845; \citealt{2011ApJS..192....3P,2013ApJS..208....4P,2015ApJS..220...15P}), which evolves the structure of the donor star and orbit in time, but treats the accretor as a point mass. Mass transfer is assumed to be fully non-conservative (MESA parameter ${\rm mass\_transfer\_beta}=1$), so the accretor mass $M_{\rm a,i}$ is a constant in time and mass-loss from the binary is assumed to take place in a fast wind from the accretor. The physical basis for this assumption is that accretion disk winds may limit the mass that falls on to the accretor, and nova explosions may remove the accreted mass.

The mixing length parameter is set to $\alpha_{\rm ML} = 1.9$. The ZAMS metallicity of all stars is $Z=0.01$, which is characteristic of the disk stars in the Galaxy \citep{2014A&A...562A..71B}. As stars evolve faster for lower metallicity with the same star mass \citep{2016A&A...595A..35I}, this metallicity choice helps accelerate the production of a WD within the age of the Galaxy. The nuclear burning network used is ``pp\_and\_cno\_extras", which includes $\mathrm{^1H,\,\,^3He,\,\,^4He,\,\,^{12}C,\,\,^{14}N,\,\,^{16}O,\,\,^{20}Ne,\,\,^{24}Mg}$ and extended networks which comprise the pp-chain and CNO cycle. Element diffusion is included over the entire evolution, starting from ZAMS end extending through the WD cooling track. This setting is crutial in regards to the critical mass at which H flashes occur, as well as to the number of flashes before the WD cooling as found by \cite{2016A&A...595A..35I}. The setting ``diffusion\_use\_cgs\_solver = .true." is used to allow for electron degeneracy in the diffusion physics. Five classes of elements, ${\rm ^1H,\,\,^3He,\,\,^4He,\,\,^{16}O,\,\,^{56}Fe}$, are evolved. Helium core masses, $M_{\rm c}$, reported here are computed as the mass interior to the point where the mass fraction of $^1$H is 1\% that of $^4$He.

The total orbital angular momentum, $J$, evolves through torques due to magnetic braking ($\dot{J}_{\rm mb}$), gravitational waves ($\dot{J}_{\rm gr}$, \citealt{1975ctf..book.....L}) and mass-loss from the binary ($\dot{J}_{\rm ml }$)
\begin{equation}
\dot{J}=\dot{J}_{\text{mb}}+\dot{J}_{\text{gr}}+\dot{J}_{\text{ml}}.
\label{Jdot_tot}
\end{equation}
Angular momentum loss from the binary due to a fast wind in the viscinity of the accretor is \citep{2006csxs.book..623T}
\begin{equation}
\dot{J}_{\text{ml}} = J\ \frac{M_{\rm d} \dot{M}_{\rm d}}{M_{\rm a}(M_{\rm d}+M_{\rm a})}.
\label{Jdot_ml}
\end{equation}
The mass-loss torque $\dot{J}_{\text{ml}}$ is important during the thermal timescale mass transfer TTMT, when $M_{\rm d} \ga M_{\rm a}$ and the mass-loss rate of the donor $\dot{M}_{\rm d}$ is high. Thereafter, $\dot{J}_{\text{mb}}$ takes over until the convection zone thins. The gravitational wave torque $\dot{J}_{\text{gr}}$ is important for short periods $P_{\rm orb} \la 3\, \rm hr$, and is the dominant torque at the second phase of MT at $P_{\rm orb} \la 1\, \rm hr$.

For thick convection zones with mass fraction $q_{\rm conv}>0.02$, the magnetic braking formula of \citet{1983ApJ...275..713R} is used. MESA's implementation is to set $\dot{J}_{\rm mb}=0$ when the fraction of mass in the convection zone $q_{\rm conv} > 0.75$ to implement the above-mentioned reduction at small stellar mass. To take into account reduced magnetic braking when the surface convection zone is thin, the ansatz from \citet{2002ApJ...565.1107P} is that $\dot{J}_{\rm mb}$ is reduced by an exponential factor as the convection zone mass becomes small. The end result used in the simulations is then
\begin{eqnarray}
\dot{J}_{\rm mb} & = & -3.8 \times 10^{-30}M_{\rm d} R_{\odot}^4 \bigg( \frac{R_{\rm d}}{R_{\odot}} \bigg)^{\gamma} \omega^3
\nonumber \\  & \times &  
\left\{ 
\begin{tabular}{rc}
0, & $ 1 > q_{\rm conv} > 0.75$ \\
1, & $0.75 > q_{\rm conv} > 0.02$ \\
$e^{1-0.02/q_{\rm conv}}$, & $q_{\rm conv} < 0.02$
\end{tabular}
\right.
\label{Jdot_mb}
\end{eqnarray}
where $\dot{J}_{\rm mb}$ is in CGS units ${\rm g\,cm^2\,s^{-1}}$, magnetic braking index $\gamma=4$ was used in the calculations, $\omega=2\pi/P_{\rm orb}$ is the orbital angular velocity in ${\rm rad\,s^{-1}}$, $R_{\rm d}$ is the donor star radius. The mass fraction $q_{\rm conv}=0.02$ is for the current solar convection zone and so magnetic braking is suppressed on the MS for more massive stars. Because the donors are evolved, their radii shrink less than for unevolved donors and there is only a weak dependence on $\gamma$. Calculations with $\gamma=3$ and $4$ gave similar results. 

A side effect of the reduced $\dot{J}_{\rm mb}$ at small $q_{\rm conv}$ is that donors with mass $1.3\la M_{\rm d}/M_{\odot} \la 1.4$, which have small magnetic braking on the MS, can have sufficient magnetic braking as evolved donors, with thicker convection zones that they work well as the progenitors of ELM WDs. Their MS lifetime is much shorter than a $0.9\leqslant M_{\rm d}/M_{\odot} \leqslant 1.1$ donor, so this leaves more time for the WD to cool to small $T_{\rm eff} \la 9000\, \rm K$ and enter the blue edge of the instability strip. Simulations of donors with larger masses $M_{\rm d} \ga 1.5\, M_\odot$ had difficulty forming an ELM WD because $M_{\rm c} \ga 0.1\, M_\odot$ at the end of the MS which, when combined with core growth during the accretion phase, makes them too large to be the ELM WD with $M<0.18\, M_\odot$. Furthermore, the orbits are much wider than for the $1.0\leqslant M_{\rm d}/M_{\odot} \leqslant 1.4$ cases.

Simulations in which the MT rose sharply and exceeded $|\dot{M}_{\rm d}|> 10^{-3}\, M_\odot\ {\rm yr}^{-1}$ were stopped and labeled as exhibiting unstable MT. This occurs if the initial mass ratio $q_{\rm i}=M_{\rm d,i}/M_{\rm a,i}$ is too large, where $M_{\rm d,i}$ is the initial donor star mass, $M_{\rm a,i}$ is the initial accretor mass, and is exacerbated by wider orbital separations such that the donor was well up the giant branch when MT commenced. As discussed in Appendix \ref{sec:ce}, unstable MT and CE may lead to merging for the $M_{\rm c} \la 0.1\, M_\odot$ here. 

In our model, the ELM WD progenitor is assumed to be the initially less massive star, and the initially more massive star becomes the ELM WD companion, itself a WD. The initially more massive star is assumed to form a WD through a CE phase, because short orbital periods from 1 to 3 days are required in the second phase of MT to form the ELM WD. Let $M_{\rm 1}$ be the mass of the initially more more massive star, $M_{\rm 2}$ the mass of the initially less massive star, and $a_{\rm CE,i}$ the initial semi-major axis before the CE. Notice that the subscript ``1", ``2" and ``CE" are only used in this section, and indicate the star parameters before the stable RLOF phase. For a wide initial orbit, a core mass $M_{\rm 1,c}$ is formed in star 1, and by removing the envelope, $M_{\rm 1,c}$ is the mass of the ELM companion. Applying the CE energy equation (Equation \ref{eq:ce_energy} in the appendix), and expressing the answer in terms of the post-CE (but pre-magnetic braking) orbital period $P_{\rm CE,orb,f}$, gives
\be
&& \left( \frac{G(M_{\rm 1,c}+M_{\rm 2})P_{\rm CE,orb,f}^2}{4\pi^2} \right)^{1/3}  = R_{\rm 1}(M_{\rm 1,c}) 
\nonumber \\ & \times & \left( \frac{M_{\rm 1,c}M_{\rm 2}}{ (2/\alpha \lambda) M_{\rm 1}(M_{\rm 1}-M_{\rm 1,c}) + M_{\rm 1} M_{\rm 2} r_{\rm L}(M_{\rm 1}/M_{\rm 2})} \right).
\label{eq:companion_mass}
\ee
with the appropriate $R_1(M_{\rm 1,c})$ relation for each core mass range, this equation can be solved for $P_{\rm CE,orb,f}$ during the evolution, where $P_{\rm CE,orb,f}$ is the post-common envelope (but pre-magnetic braking) orbital period, $\lambda \simeq 1$ is a mass-dependent factor describing the binding energy, $\alpha$ is the efficiency of tapping orbital energy to remove the envelope, and $r_{\rm L}a$ is the effective radius of the Roche lobe, and $r_{\rm L}$ is a parameter defined in \cite{1983ApJ...268..368E}. MESA models for $M_{\rm 1,i}=$1, 2, 3, 4, 5 $M_{\odot}$ were used to find $R_{\rm 1}$, $M_{\rm 1,c}$ and $M_{\rm 1}$ during the evolution. The radius grows non-monotonically, so this leads to gaps in $M_{\rm 1,c}$ over regions where the radius decreases below its maximum value.

\begin{figure}[]
\centering
\includegraphics[width=0.5\textwidth]{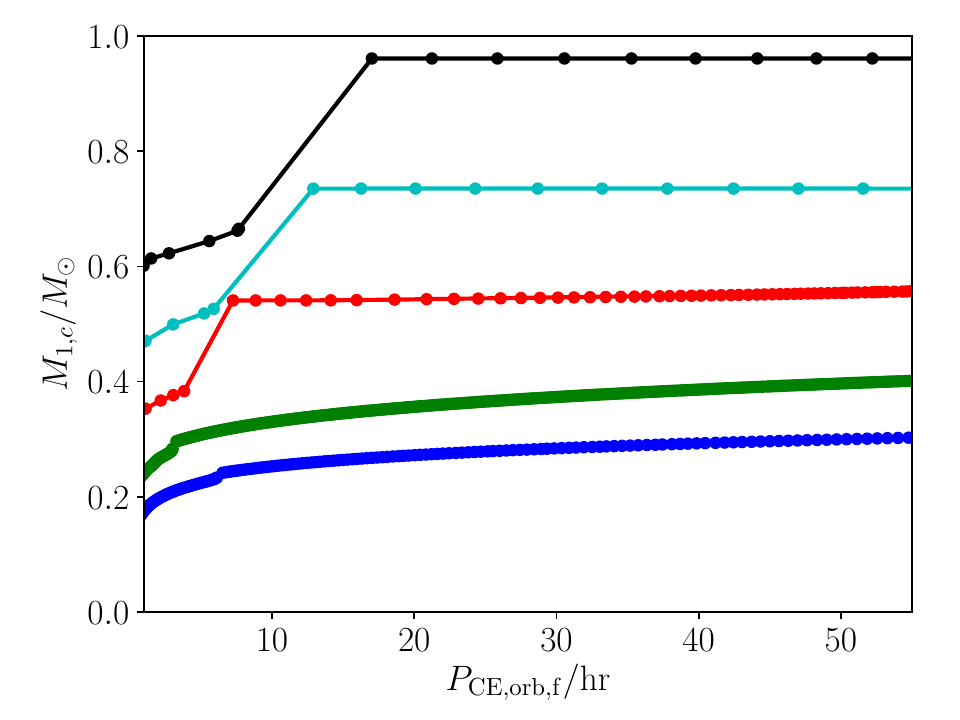}
\caption{ ELM WD companion mass $M_{\rm 1,c}$, as a function of post-common envelope (but pre-magnetic braking) orbital period, $P_{\rm CE,orb,f}$. The lines represent different progenitor mass $M_1/M_\odot=1,2,3,4,5$ for the companion. The ELM WD progenitor is assumed to have mass $M_2=1.3\, M_\odot$.}
\label{fig:companion_mass}
\end{figure}

Figure \ref{fig:companion_mass} shows numerical solutions of Equation \ref{eq:companion_mass} for companion mass $M_{1,c}$ as a function of $P_{\rm CE,orb,f}$. The ELM WD progenitor mass has been fixed at $M_2=1.3\, M_\odot$, and five different $M_1$ have been used to give the different lines. The product $\alpha \lambda$ is set to 2 for convenience. There is a general trend that $M_{\rm 1,c}$ must be larger for larger orbital period or $M_1$, in order that the orbital energy release can balance the binding energy. During the second phase of MT, the companion is the accretor and so $M_{\rm a,i}=M_{\rm 1,c}$, and the progenitor of the ELM WD is the donor, so $M_{\rm d,i}=M_2$. 

The separation at the onset of the RLOF should be slightly greater than 5 $R_{\odot}$ to form an ELM WD. If $M_{\rm 1,c}$ is fixed at 0.6 $M_{\odot}$ with $M_2=1.3\, M_\odot$ and the separation after the CE $a_{\rm CE,f}=5 R_{\odot}$, there are still two free (but not completely free) parameters $M_1$ and the orbital period before the CE $P_{\rm CE,i}$. Moreover, $M_1$ is greater than $M_2$ because the massive star evolves first. This can lead to a CE phase. And $M_1/M_2$ is greater than one to have unstable MT followed by a CE phase \citep{2011ApJ...739L..48W}. For $M_1= 2\,\,M_{\odot}$, $P_{\rm CE,i}$ is 7.6 days.

\section{ results on binary formation}
\label{sec:results}

\subsection{ The Fiducial Case }
\label{Binary Formation}
\begin{figure}[tp]
\centering
\includegraphics[width=0.55\textwidth]{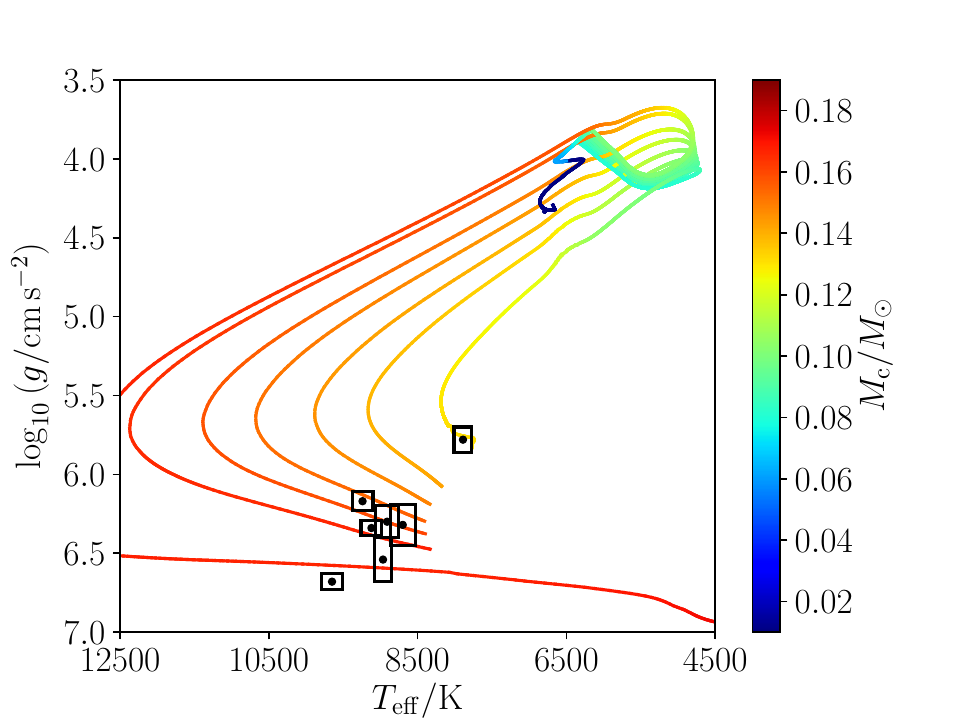}
\caption{Evolutionary models for the donor of initial mass $M_{\rm d,i}=1.3\, M_{\odot}$ and (constant) accretor mass $M_{\rm a,i}=0.6\, M_{\odot}$. The figure shows the entire range of ELM WDs, which is covered by the range of initial orbital periods $P_{\rm orb,i}=$0.90, 0.93, 0.95, 0.97, 0.99, 1.02 days, from right to left. In addition, a model with slightly larger $P_{\rm orb,i}=1.03\,{\rm day}$ is shown, for which shell flashes occur on the WD cooling track. The color indicates the helium core mass, $M_{\rm c}\,(M_\odot)$. The black points with error bars are the seven pulsating ELM WDs. The track with $P_{\rm orb,i}=0.90$ day gives the minimum mass of the ELM WD to be $M_{\rm d,f}=0.146 M_{\odot}$. The $P_{\rm orb,i}=1.03\, {\rm day}$ model yields a WD of mass $M_{\rm d,f}=0.179\, M_{\odot}$. The evolution between the first and last shell flashes is not shown on the plot, for clarity.}
\label{fig:many tracks from RLOF}
\vspace{6pt}
\end{figure}

Figure \ref{fig:many tracks from RLOF} displays evolution tracks in the $\log_{10}\,g$ versus $T_{ \rm eff}$ plane for the fiducial case with $M_{\rm d,i}=1.3\, M_\odot$ and (constant) accretor mass $M_{\rm a,i}=0.6\, M_\odot$. The entire range of ELM WDs is covered by the initial orbital period $P_{\rm orb,i}=$0.90, 0.91, 0.93, 0.95, 0.97, 0.99, 1.02 days, after the CE but before the RLOF phase. The narrow range of $P_{\rm orb,i}$ which produce ELM WD is similar to the result found by \cite{2017MNRAS.464..237S}. The donor in the track with $P_{\rm orb,i}=1.03\, \rm day$ has a core mass large enough that diffusion-aided shell flashes occur on the WD cooling track. The color indicates $M_{\rm c}$. The black points with error boxes represent the seven pulsating ELM WDs with parameters derived using 3-D atmosphere models \citep{2015ApJ...809..148T} except J1618+3854 (since only the $\log_{10}\,g$ and $T_{ \rm eff}$ from 1-D atmosphere model is given in other references), with the half width of the box showing the measurement uncertainty.

All runs begin at the ZAMS with $\log_{10}\,(g/{\rm cm\,s^{-2}}) = 4.4$ and $T_{ \rm eff}=6500\,{\rm K}$. Along the MS, and as the star evolves to the RGB, its radius increases with $M_{\rm c}$, and so wider orbits come into contact with larger $M_{\rm c}$. The ELM WD commences MT with $0.06\la M_{\rm c}/M_{\odot} \la 0.1$, and $M_{\rm c}$ increases during the MT phase. Figure \ref{fig:many tracks from RLOF} shows that models with larger $M_{\rm c}$ evolve to a higher maximum $T_{\rm eff}$, the elbow in the curve that separates the pre-WD phase (increasing $T_{\rm eff}$) from the WD cooling track (decreasing $T_{\rm eff}$). This plot shows the same behavior between shell flashes, that the loops in the $\log g-T_{\rm eff}$ plane become larger, evolving to higher maximum $T_{\rm eff}$. As a result, when systems with shell flashes enter the WD cooling track, their evolution is more nearly horizontal, at constant $\log_{10}\,g$. This gives rise to a wedge in the $\log_{10}\,g-T_{\rm eff}$ plane which separates the ELM WD with $M \la 0.18\, M_\odot$ without shell flashes from the slightly more massive WDs, with $M \ga 0.18\, M_\odot$, which do have shell flashes. The hydrogen-rich envelope is thinner after the shell flashes, so the residual hydrogen burning is smaller and the system evolves to lower $T_{\rm eff}$ more quickly. All runs were evolved to an age 13.7 Gyr, except the one run in the Figure which come back into contact. Furthermore, the low-mass donor star evolves slower and cannot reach the WD cooling phase by the age of the Galactic disk (10 Gyr). We extended the evolution to 13.7 Gyr to see if the WD cooling phase can be reached within a Hubble time. The ELM WDs in Figure \ref{fig:many tracks from RLOF} have much longer cooling times, and only get down to $T_{\rm eff} \simeq 8000\, \rm K$, while the run with shell flashes in the lower panel makes it down to $T_{\rm eff} \la 4000\, \rm K$. The observed systems evidently span the range of ELM WDs with thick envelopes as well as those which have undergone shell flashes.

For smaller $P_{\rm orb,i}$, the donor comes into contact at core mass $0.01\la M_{\rm c}/M_{\odot} \la 0.07$, 
and stays in contact to short $P_{\rm orb} \la 1\rm hr$. For systems that come into contact early on the MS, at very small $M_{\rm c} \la 0.01\, M_\odot$, standard CV evolution with a period gap at 2-3 hours is recovered. However, the radiative core is small or nonexistent in this case, and they are not expected to be g-mode pulsators. The core mass at contact for these cases is small, at roughly $M_{\rm c} \la 0.06\, M_\odot$, in agreement with \citet{2003MNRAS.340.1214P}.

\begin{figure}[tp]
\centering
\includegraphics[width=0.5\textwidth]{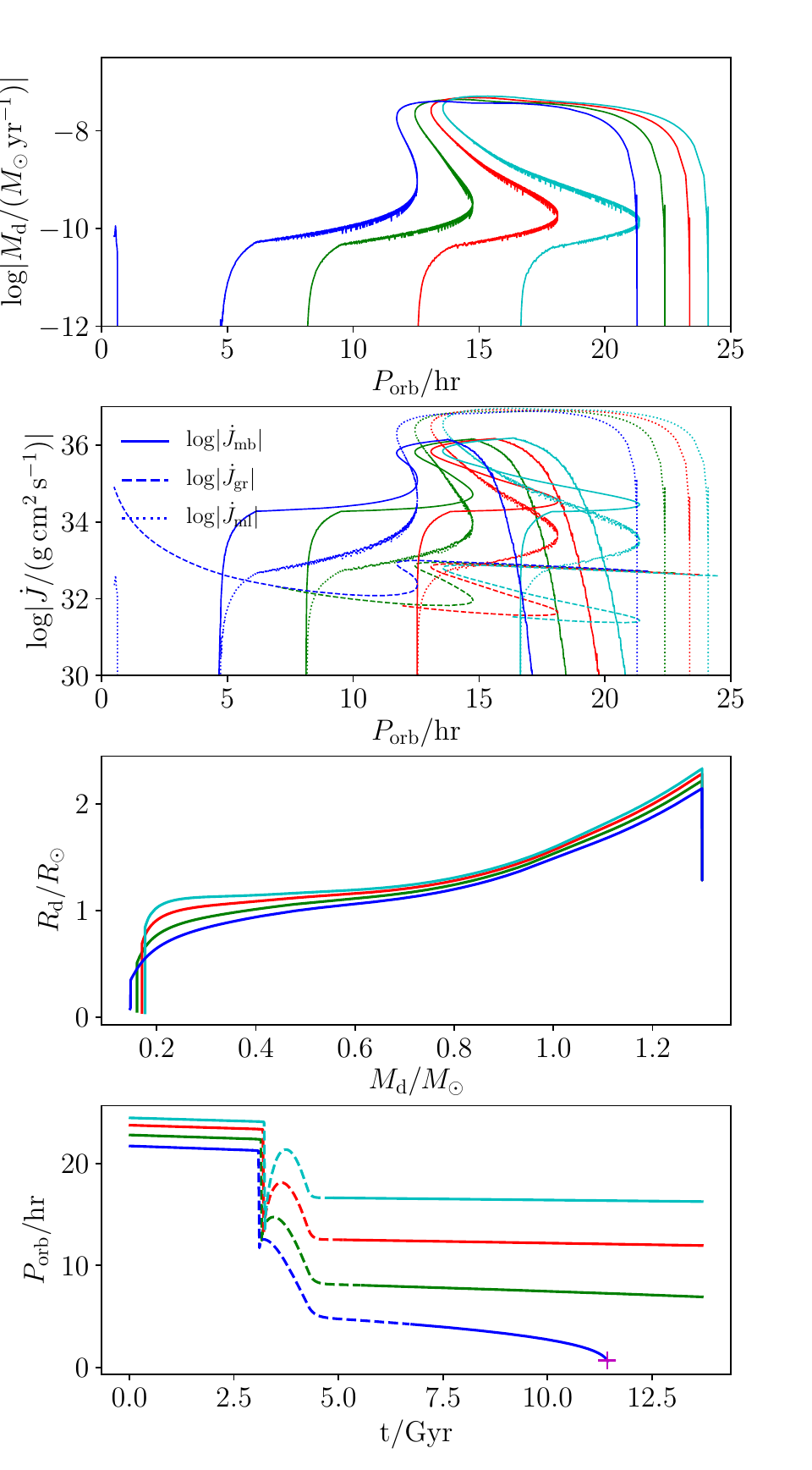}
\caption{Several evolutionary tracks with $P_{\rm orb,i}=$ 0.90 (blue), 0.95 (green), 0.99 (red) and 1.02 (cyan) day for $M_{\rm d,i}= 1.3\, M_{\odot}$ and $M_{\rm a,i}=0.6\, M_{\odot}$ (see Figure \ref{fig:many tracks from RLOF}). From top to bottom, the panels give the mass-loss rate of the donor star ($\dot{M}_{\rm d}$) versus orbital period ($P_{\rm orb}$), the separate contributions to the orbital angular momentum loss rate ($\dot{J}_{\rm ml}$, $\dot{J}_{\rm mb}$ and $\dot{J}_{\rm gr}$) versus $P_{\rm orb}$, the donor star radius ($R_{\rm d}$) versus donor mass ($M_{\rm d}$), and $P_{\rm orb}$ versus age (t). In the bottom panel, on the $P_{\rm orb,i}=0.90$ day track, the magenta cross marks the beginning of the second phase of MT.}
\label{tracks13_060.pdf}
\end{figure}

Figure \ref{tracks13_060.pdf} shows $\dot{M}_{\rm d}$ versus $P_{\rm orb}$ (top panel), $\dot{J}$ contributions versus $P_{\rm orb}$ (second panel), donor $R_{\rm d}$ versus $M_{\rm d}$ (third panel) and $P_{\rm orb}$ versus age (bottom panel). The initial periods are $P_{\rm orb,i}=0.90$ (blue), 0.95 (green), 0.99 (red) and 1.02 days (cyan). 

In the top two panels, evolutionary tracks producing ELM WDs start at long periods and proceed to shorter periods on the whole. Magnetic braking is small for $M_{\rm d,i}=1.3\, M_\odot$ on the MS, due to the small surface convection zone, so $P_{\rm orb}$ is nearly constant during that time. When the system first comes into contact, TTMT results in high mass-loss rates $10^{-8} \la \dot{M}_{\rm d}/ M_\odot \, {\rm yr^{-1}} \la 10^{-7}$. TTMT continues until the ratio $M_{\rm d}/M_{\rm a}$ decreases to the critical value (1 for conservative transfer, see \citealt{2012ApJ...744...12W}) at which point TTMT ends, and the much slower nuclear or $\dot{J}$ timescale MT takes over. During TTMT, $\dot{J}_{\rm ml}$ dominates, due to the high accretion rates (second panel). Shortly thereafter, the increased size of the convection zone removes the suppression of $\dot{J}_{\rm mb}$, and it subsequently dominates until MT turns off at $5 \leqslant P_{\rm orb}/{\rm hr} \leqslant 11$. Subsequently $J_{\rm gr}$ dominates and all systems undergo orbital decay. Only the lowest mass, less evolved donors undergo sufficient orbital decay to come back into contact at $P_{\rm orb} \approx 1\, \rm hr$. For larger $M_{\rm c}$, evolution driven by the expansion of the star as it tries to ascend the RGB becomes more important than orbital shrinkage due to magnetic braking, leading to a period bifurcation separating the orbits which shrink from those which expand \citep{2003MNRAS.340.1214P}.

The third panel in Figure \ref{tracks13_060.pdf} shows donor radius as a function of mass during the evolution. Before contact, $R_{\rm d}$ increases with $M_{\rm c}$. Smaller $P_{\rm orb,i}$ runs commence MT first, at smaller $R_{\rm d}$, while larger $P_{\rm orb,i}$ allows $R_{\rm d}$ to grow further. During TTMT, the high MT rate causes the radius to be slightly inflated. As discussed in Appendix \ref{sec:conservative} (see Figure \ref{fig:R_vs_M}), the decrease in mass of the hydrogen-rich envelope leads to smaller hydrogen shell-burning luminosity, accompanied by shrinkage of the radius. This is seen in the steep drop in radius in third panel of Figure \ref{tracks13_060.pdf}, leading the system to fall out of contact, as shown in the first panel. Models with thick hydrogen envelopes have modest shrinkage in radius beyond that point. The lowest mass model comes back into contact, evolving toward smaller $M_{\rm d}$.

The bottom panel in Figure \ref{tracks13_060.pdf} shows $P_{\rm orb}$ versus age. The evolution starts on the left, with tracks at different $P_{\rm orb,i}$ denoted by solid lines. When MT commences (dashed lines starting at 3 Gyr), rapid orbital decay occurs during TTMT. Then the slower orbit evolution on the $\dot{J}$ timescale lasts 2 to 3 Gyr. During this slow phase of MT, slight orbit expansion occurs for large $M_{\rm c}$, while continued orbital decay occurs for small $M_{\rm c}$. When MT ceases, the donor becomes an ELM WD near 5 to 6 Gyr (solid lines). The three largest $P_{\rm orb,i}$ and $M_{\rm c}$ tracks show only modest orbital decay due to $\dot{J}_{\rm gr}$, while the lowest line decays from $P_{\rm orb}=5\, \rm hr$ to 1 hr, at which point MT re-commences (the magenta cross).

\begin{figure}[tp]
\centering
\includegraphics[width=0.5\textwidth]{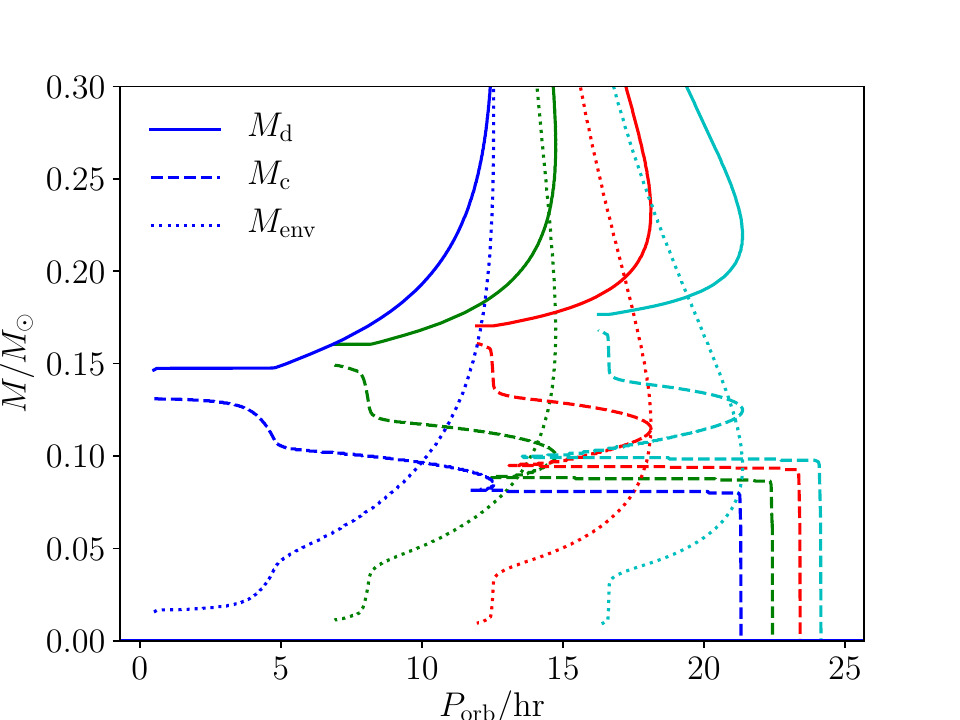}
\caption{Donor star mass ($M_{\rm d}$, solid), helium core mass ($M_{\rm c}$, dashed) and envelope mass ($M_{\rm env}=M_{\rm d}-M_{\rm c}$, dotted) as a function of $P_{\rm orb}$, for $P_{\rm orb,i}=$ 0.90 (blue), 0.95 (green), 0.99 (red) and 1.02 (cyan) days, $M_{\rm d,i}=1.3\,M_{\odot}$ and $M_{\rm a,i}=0.6\,M_{\odot}$.} 
\label{Masses13_Porb.pdf}
\end{figure}

Figure \ref{Masses13_Porb.pdf} shows total donor mass $M_{\rm d}$, helium core mass $M_{\rm c}$ and envelope mass $M_{\rm env}=M_{\rm d}-M_{\rm c}$ versus $P_{\rm orb}$. The tracks start from the right of the plot near $20 \la P_{\rm orb}/{\rm hr} \la 25$. The total donor mass $M_{\rm d}$ decreases downward during MT, and becomes constant when MT ceases. The envelope mass is seen to smoothly decrease, until the end of MT at $5 \leqslant P_{\rm orb}/{\rm hr} \leqslant 20$. The envelope mass $M_{\rm env}$ continues to decrease due to residual hydrogen burning, while the orbit slowly decays due to $\dot{J}_{\rm gr}$. The $M_{\rm c}$ lines initially rise vertically, as $M_{\rm c}$ increases before MT. The TTMT phase is so short that there is no time for $M_{\rm c}$ to grow above $0.07 \la M_{\rm c}/M_\odot \la 0.10$. The nuclear timescale MT is much longer, and $M_{\rm c}$ increases to $0.13 \la M_{\rm c}/M_\odot \la 0.15$. After MT, an additional $0.02 \la M_{\rm c}/M_\odot \la 0.03$ is converted from envelope to core by nuclear burning, during which time the orbit decays due to $\dot{J}_{\rm gr}$.

\begin{figure}[tp]
\centering
\includegraphics[width=0.5\textwidth]{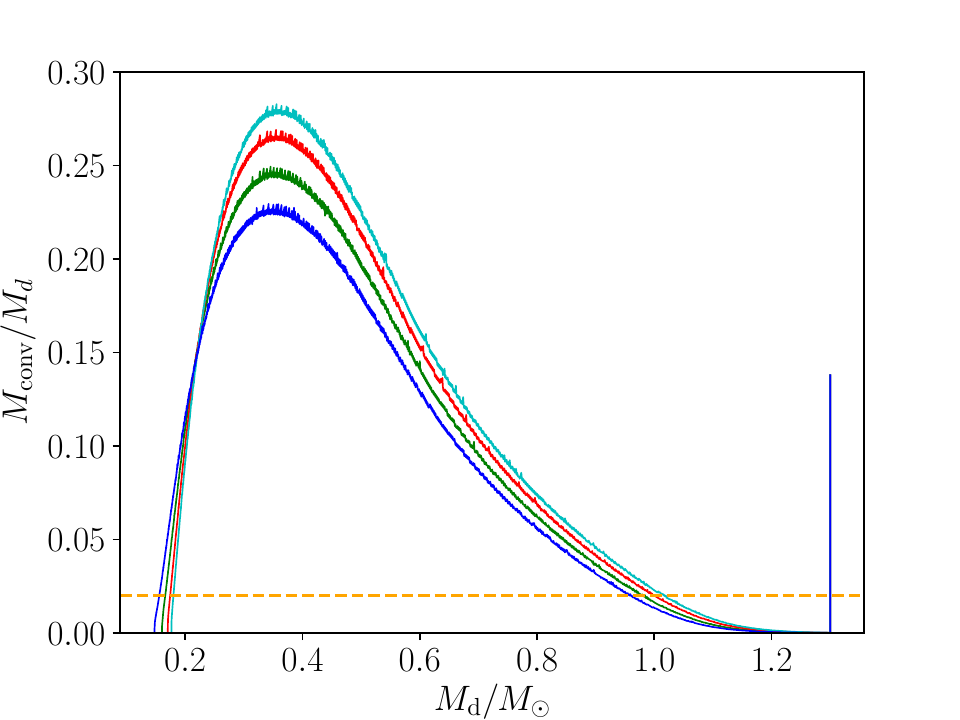}
\caption{Mass fraction of the convective zone ($M_{\rm conv}/M_{\rm d}$) versus $M_{\rm d}$ for $M_{\rm d,i} = 1.3 M_{\odot}$, $M_{\rm a,i} = 0.6 M_{\odot}$ and $P_{\rm orb,i}=$ 0.9 (blue), 0.95 (green), 0.99 (red), 1.02(cyan) days. The horizontal orange dashed line marks the value of $M_{\rm conv}/M_{\rm d}$ above which the full magnetic braking is applied, and below which magnetic braking is suppressed (see Equation \ref{Jdot_mb}).} 
\label{q_vs_mass.pdf}
\end{figure}

From the discussion in Section \ref{sec:elm_from_mb} (see Equation \ref{Jdot_mb}), the thickness of the convective envelope is an important parameter for the effectiveness of magnetic braking. Figure \ref{q_vs_mass.pdf} shows the mass of the convective envelope $M_{\rm conv}$ as a function of $M_{\rm d}$. The systems evolve from right to left during MT. The $M_{\rm d,i}=1.3\, M_\odot$ donor has a small convective envelope on the MS. The spike at $M_{\rm d}=1.3\,M_{\odot}$, is due to the convective core on the MS. As $M_{\rm c}$ grows and the shell burning luminosity increases, $M_{\rm conv}$ increases. When nearly all the hydrogen-rich envelope has been lost, the luminosity drops and the convection zone shrinks again. The orange dashed line gives the threshold below which magnetic braking is exponentially suppressed. For the ELM WD, $M_{\rm conv}$ drops below the threshold and magnetic braking shuts off. 

\begin{figure}[tp]
\centering
\includegraphics[width=0.5\textwidth]{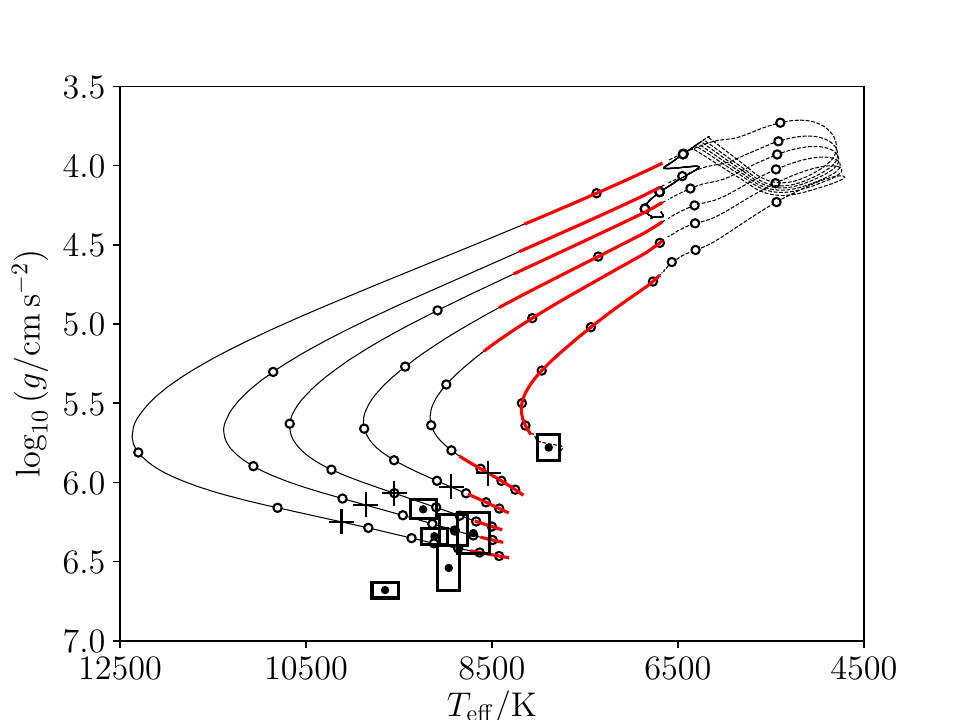}
\caption{ Gravity (${\rm log}\,g$) versus $T_{\rm eff}$ for $M_{\rm d,i} = 1.3 M_{\odot}$, $M_{\rm a,i} = 0.6 M_{\odot}$ and $P_{\rm orb,i}=$ 0.90, 0.93, 0.95, 0.97, 0.99, 1.02 days. The hollow dots are placed at 1 Gyr intervals. The solid line shows phases where the system is out of contact, while the dashed lines show phases where MT is occurring. The black crosses show the places where the model reaches the diffusive equilibrium, see Section \ref{sec:Instability Strip and Seismology} for details.}
\label{fig:logg_Teff_13_060.pdf}
\end{figure}

Figure \ref{fig:logg_Teff_13_060.pdf} shows the same ${\rm log}\,g$ versus $T_{\rm eff}$ as Figure \ref{fig:many tracks from RLOF}. The circles are placed at 1 Gyr intervals. Systems covered by the tracks have ages 9-12 Gyrs. The solid line shows phases where the system is out of contact, while the dashed lines show phases where MT is occurring. Following \citet{2010ApJ...718..441S}, phases for which the lowest order $\ell=1$ g-mode is unstable are estimated by Brickhill's criterion, $P(\rm g1) \leqslant 8\pi t_{\rm th}$, and are covered by red lines, where $P(\rm g1)$ is the mode period of the lowest order g-mode and $t_{\rm th}$ is the thermal time at the base of the surface convection zone. The two data points at high ${\rm log}\,g$ would require $M_{\rm d,f}>0.18\,M_{\odot}$ tracks which are not shown on the plot. The estimate of the instability strip used here appears to give too cool a blue edge $T_{\rm eff}$ to explain the systems near $8500 \leqslant T_{\rm eff}/{\rm K} \leqslant 9,500$. 

\begin{figure}[tp]
\centering
\includegraphics[width=0.5\textwidth]{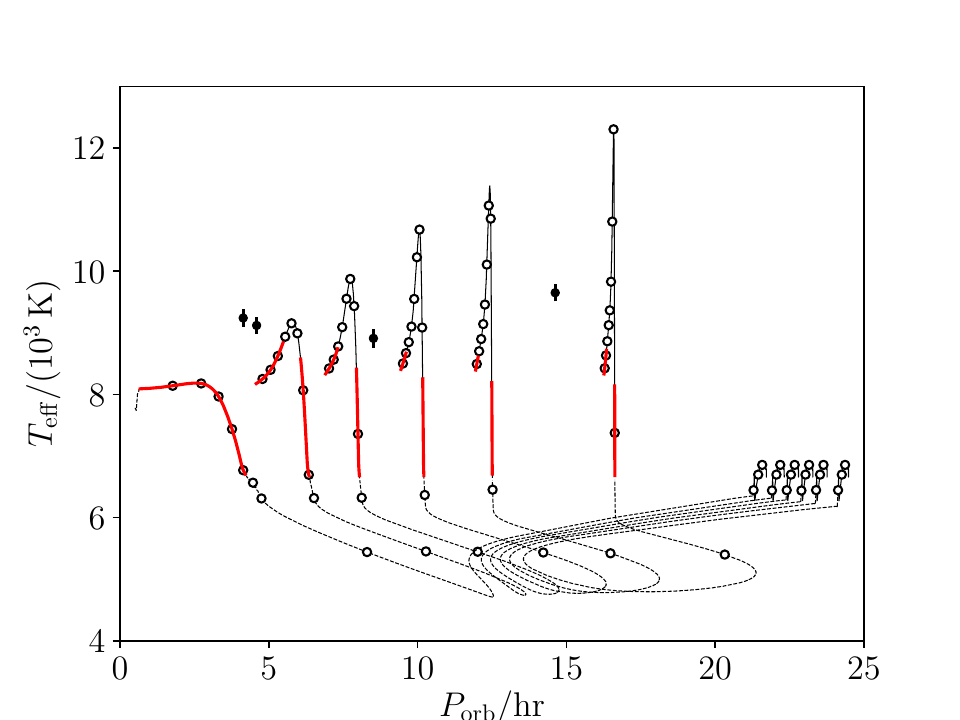}
\caption{ Effective temperature ($T_{\rm eff}$) versus $P_{\rm orb}$ for $M_{\rm d,i} = 1.3 M_{\odot}$, $M_{\rm a,i} = 0.6 M_{\odot}$ and $P_{\rm orb,i}=$ 0.90, 0.93, 0.95, 0.97, 0.99, 1.02 days. The solid line shows phases where the system is out of contact, while the dashed lines show phases where MT is occurring. The red lines show models for which the g1 mode is unstable by Brickhill's criterion.}
\label{Teff_P_13_060.pdf}
\end{figure}

Figure \ref{Teff_P_13_060.pdf} displays $T_{\rm eff}$ versus $P_{\rm orb}$ for $P_{\rm orb,i}=$ 0.90, 0.93, 0.95, 0.97, 0.99, 1.02 days as well as the four observed systems with measured $P_{\rm orb}$. The evolution starts from the right hand side of the plot. The dashed, solid and red lines have the same meaning as in Figure \ref{fig:logg_Teff_13_060.pdf}. The pre-WD phase starts at $T_{\rm eff} \ga 6500$ K. After the turning point as the $T_{\rm eff}$ reaches the maximum, the WD enters its cooling phase. The two pulsators with the shortest $P_{\rm orb}$ have slightly higher $T_{\rm eff}$ than the models. The lines covered by red segments show the unstable g1 mode with Brickhill's criterion. In the following discussion section it will be shown that lower-mass donor and high-mass accretor can give better agreement.

\subsection{ A More Massive Accretor}

The runs in this section use the same $M_{\rm d,i} = 1.3\,M_{\odot}$ but a heavier accretor mass $M_{\rm a,i} = 0.9\,M_{\odot}$. The structures and the evolutionary tracks of the donor stars do not change significantly with companion mass, while the orbital period of the system can be different \citep{2016A&A...595A..35I}.

\begin{figure}[tp]
\centering
\includegraphics[width=0.5\textwidth]{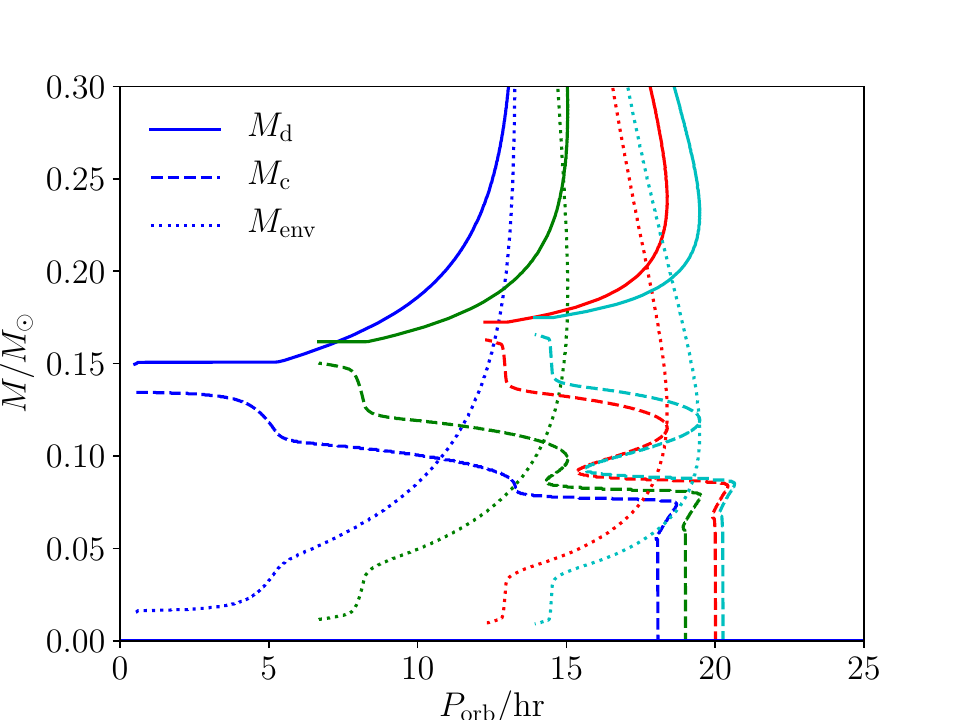}
\caption{ Same as Figure \ref{Masses13_Porb.pdf}, but for donor mass $M_{\rm d,i}=1.3\,M_{\odot}$ and heavier accretor mass $M_{\rm a,i}=0.9\,M_{\odot}$. The orbital period for each line is $P_{\rm orb,i}=$ 0.90 (blue), 0.95 (green), 0.99 (red) and 1.01 (cyan) days. }
\label{fig:Masses_P_13_090.pdf}
\end{figure}

Figure \ref{fig:Masses_P_13_090.pdf} shows {$M_{\rm d}$, $M_{\rm c}$ and $M_{\rm env}$ versus $P_{\rm orb}$, and should be compared to the fiducial case in Figure \ref{Masses13_Porb.pdf}. After the onset of RLOF, the orbit first goes outward slightly, and then shrinks until $M_{\rm d}\approx M_{\rm a}$. The accretor is larger, so the orbit does not shrink as much as the fiducial case and the mass-loss rate is also smaller. The donor star is slightly less evolved at the beginning of the RLOF because, for the same separation, the larger accretor makes the Roche-lobe radius smaller. As a result, smaller $P_{\rm orb,i}$ must be used to get the same $M_{\rm d,f}$.

\begin{figure}[tp]
\centering
\includegraphics[width=0.5\textwidth]{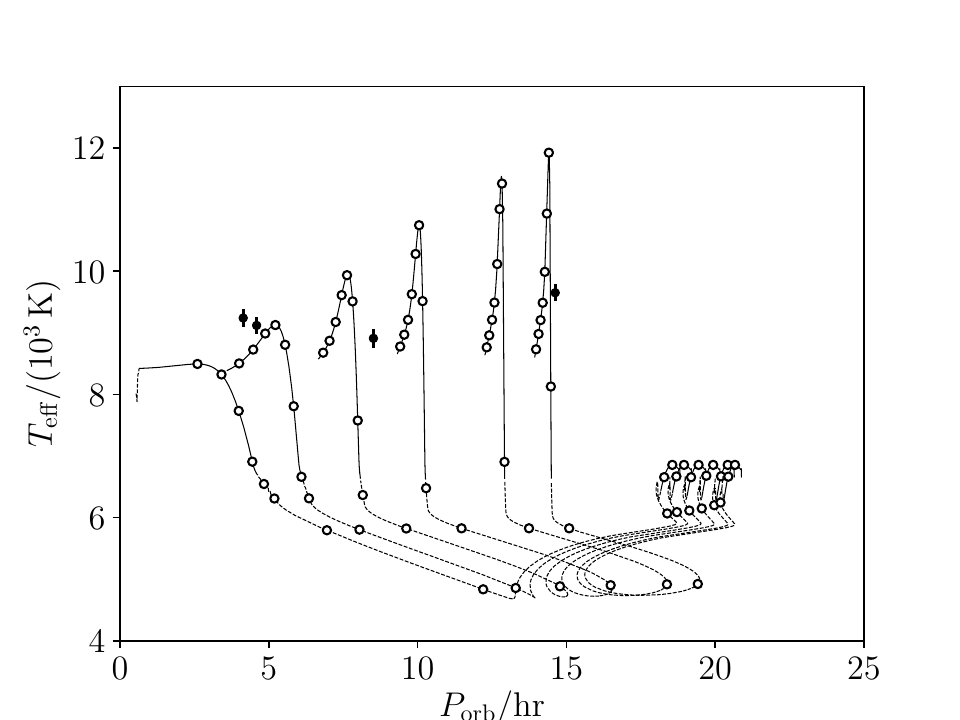}
\caption{Same with Figure \ref{Teff_P_13_060.pdf}, but for donor mass $M_{\rm d,i}=1.3\,M_{\odot}$ and larger accretor mass $M_{\rm a,i}=0.9\,M_{\odot}$. he orbital period for each line is $P_{\rm orb,i}=$ 0.78, 0.8, 0.82, 0.84, 0.86, 0.87 day.}
\label{fig:Teff_P_13_090.pdf}
\end{figure}

The main difference between Figures \ref{fig:Teff_P_13_090.pdf} and \ref{Teff_P_13_060.pdf} occurs on the WD cooling track after maximum $T_{\rm eff}$. During this long period, the larger $M_{\rm a,i}$ increases $\dot{J}_{\rm gr}$, causing the orbit to shrink faster. This is more evident for small $P_{\rm orb}$. A specific example is given in Figure \ref{fig:Teff_P_compare.pdf}, in which the pre-WD evolution is similar but the heavier accretor causes more orbital decay on the WD cooling track.

\subsection{Solar Mass Donor and Low Mass Accretor}

This section contains a comparison of evolutionary models for $M_{\rm d,i}=1.0\,M_{\odot}$ and $M_{\rm a,i}=0.45\,M_{\odot}$ to the fiducial case results in Section \ref{Binary Formation}. The companion mass is near the upper end of the mass range for helium core WDs. In addition, $M_{\rm a,i}$ is also low enough that long $P_{\rm orb,i}$ models exhibit unstable MT. Even lower $M_{\rm a,i}$ can lead to unstable MT at a broader range of $P_{\rm orb,i}$.

\begin{figure}[tp]
\centering
\includegraphics[width=0.5\textwidth]{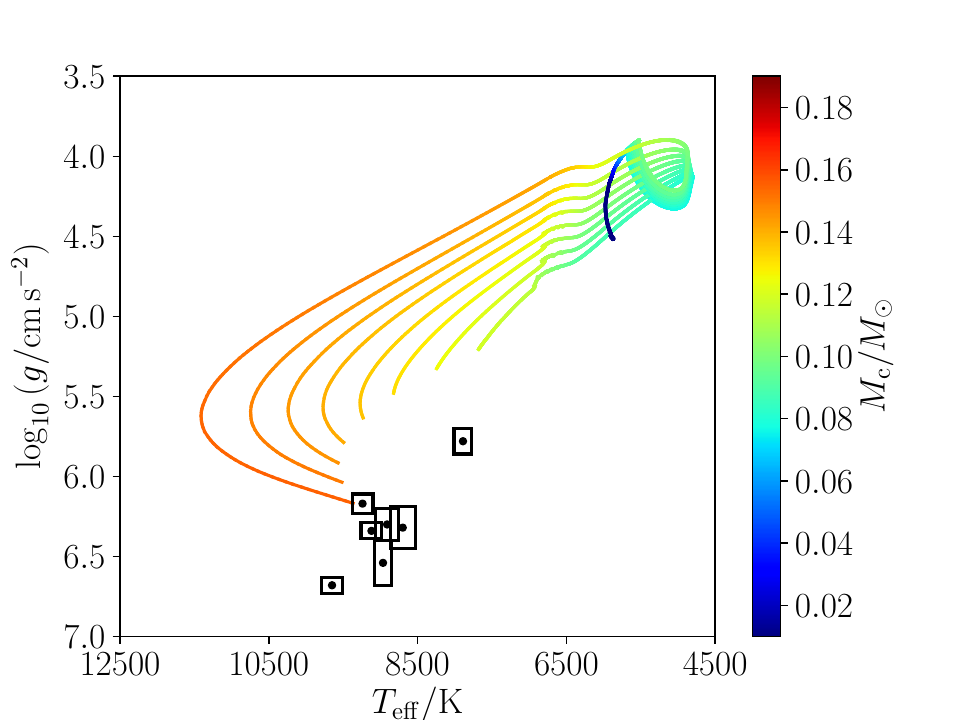}
\caption{Evolutionary models for $M_{\rm d,i}=1.0\, M_{\odot}$ and $M_{\rm a,i}=0.45\, M_{\odot}$. The figure shows the entire range of ELM WDs, which is covered by the range of initial orbital periods $P_{\rm orb,i}=$2.3, 2.35, 2.4, 2.45, 2.5, 2.55, 2.6, 2.67 days, from right to left. See Figure \ref{fig:many tracks from RLOF} for the description of the black dots.}
\label{fig:logg_Teff_color_10_045.pdf}
\end{figure}

Figure \ref{fig:logg_Teff_color_10_045.pdf} gives the evolutionary tracks for $M_{\rm d,i}=1.0\,M_{\odot}$ and $M_{\rm a,i}=0.45\,M_{\odot}$ with $P_{\rm orb,i}=$2.3, 2.35, 2.4, 2.45, 2.5, 2.55, 2.6, 2.67 days. The axes are the same as in Figure \ref{fig:many tracks from RLOF}. For $P_{\rm orb,i}<2.3$ days, accretion never ceases and the orbit shrinks to $P_{\rm orb}<1$ hour. For $P_{\rm orb,i}>2.67$ days, MT commences with a sufficiently large convective envelope that unstable MT occurs, yielding an upper limit to the WD mass produced with these evolutionary sequences. This is to be contrasted with the fiducial case in Figure \ref{fig:many tracks from RLOF}, where the ELM WD sequence joins on to the sequences of WDs at larger $P_{\rm orb,i}$ which have shell flashes. Hence the bottom track in Figure \ref{fig:many tracks from RLOF}, which shows the WD cooling track after flashes have stopped, would not occur for this case, due to the smaller $M_{\rm a,i}$ used in this section. 

Figure \ref{fig:logg_Teff_color_10_045.pdf} shows that most tracks have insufficient time to reach the $T_{\rm eff}$ of the data points. This is due to the long MS evolution. The ${\rm log}\,g$ at the elbow is slightly smaller than for the fiducial case. 

\begin{figure}[tp]
\centering
\includegraphics[width=0.5\textwidth]{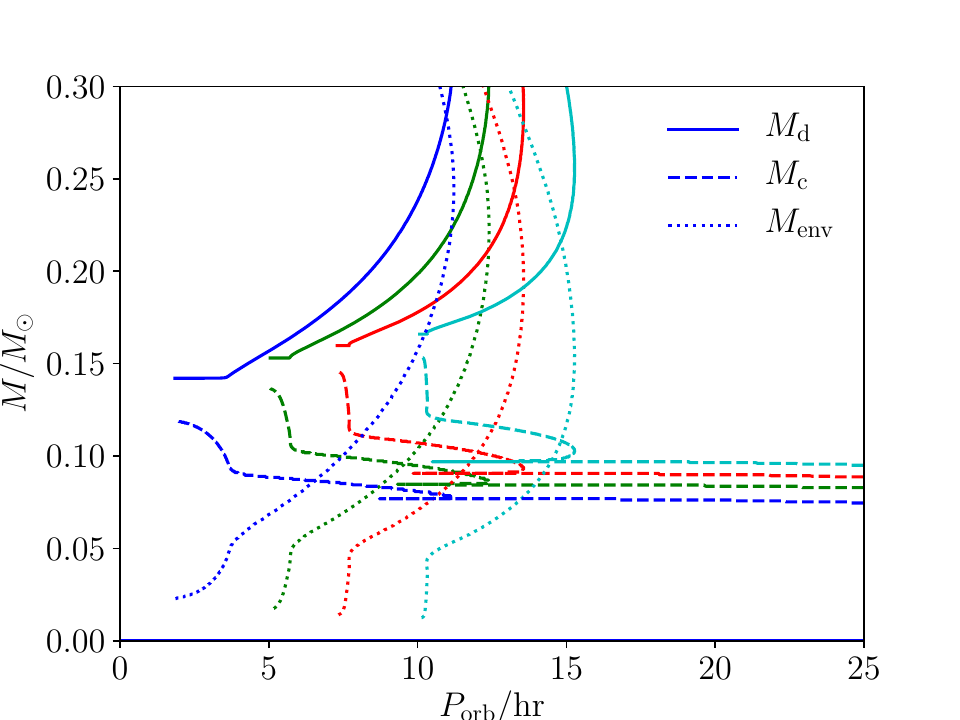}
\caption{ Same as Figure \ref{Masses13_Porb.pdf}, but for $M_{\rm d,i}=1.0\,M_{\odot}$ and a helium core accretor $M_{\rm a,i}=0.45\,M_{\odot}$, for $P_{\rm orb,i}=$ 2.3 (blue), 2.45 (green), 2.55 (red) and 2.67 (cyan) days. }
\label{fig:Masses_P_10_045zm.pdf}
\end{figure}

Similar to Figure \ref{Masses13_Porb.pdf}, Figure \ref{fig:Masses_P_10_045zm.pdf} shows $M_{\rm d}$, $M_{\rm c}$ and $M_{\rm env}$ as a function of $P_{\rm orb,i}$, now with $M_{\rm d,i}=1.0\,M_{\odot}$ and $M_{\rm a,i}=0.45\,M_{\odot}$. The selected $P_{\rm orb,i}$ are 2.3, 2.45, 2.55, 2.67 days. 
Tracks enter from the right hand side of the plot due to the large magnetic braking. This is in contrast to the fiducial case where RLOF began due to an increase in the stellar radius near the end of the MS. 
The tracks at smaller $P_{\rm orb}$ have incomplete burning of the envelope within the Hubble time.

\begin{figure}[tp]
\centering
\includegraphics[width=0.5\textwidth]{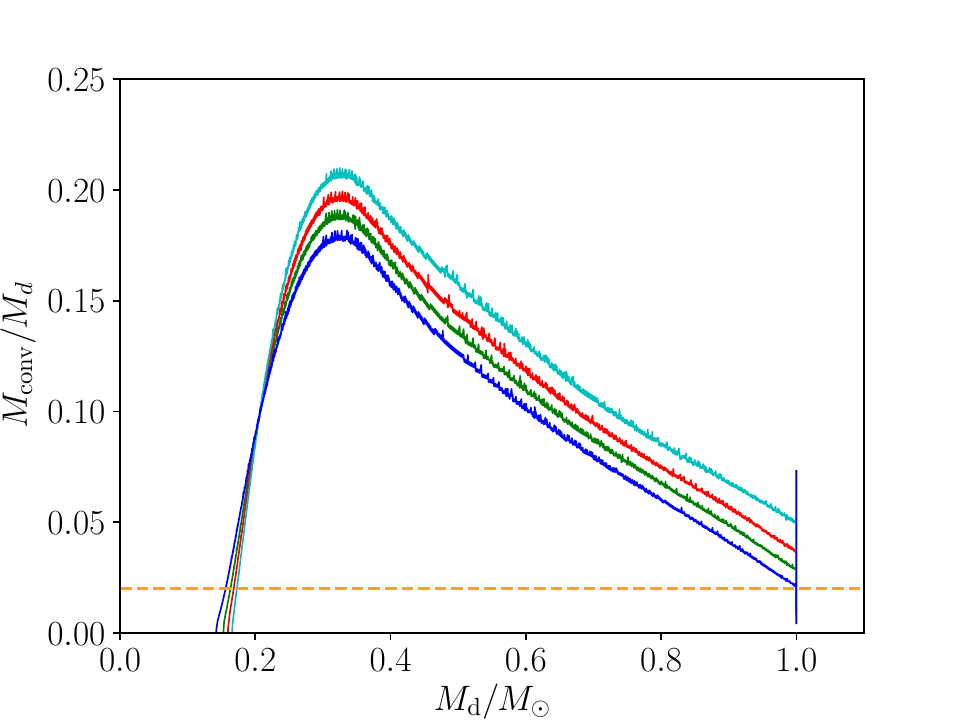}
\caption{Mass fraction of the convective zone versus donor mass $M_{\rm d}$ for $M_{\rm d,i} = 1.0 M_{\odot}$, $M_{\rm a,i} = 0.45 M_{\odot}$ and $P_{\rm orb,i}=$ 2.3 (blue), 2.45 (green), 2.55 (red), 2.67(cyan) days. See Figure \ref{q_vs_mass.pdf} to compare to the fiducial model.}
\label{fig:q_vs_m_10_045.pdf}
\end{figure}

Figure \ref{fig:q_vs_m_10_045.pdf} shows $M_{\rm conv}/M_{\rm d}$ versus $M_{\rm d}$ for $M_{\rm d,i}=1.0\,M_{\odot}$. 
The outer convection zone grows during the MS. As $P_{\rm orb,i}$ increases, the onset of RLOF occurs later, and with a larger surface convection zone. Since the outer convection zone has $q_{\rm conv}>0.02$, magnetic braking is much larger than for the fiducial case and is evident in Figure \ref{fig:Masses_P_10_045zm.pdf}. Therefore, making an ELM WD needs longer $P_{\rm orb,i}$ for $M_{\rm d,i}=1.0\,M_{\odot}$, $P_{\rm orb,i}>25$ hours. For even longer $P_{\rm orb,i}$, $M_{\rm conv}$ is sufficiently large for unstable MT to occur.

\begin{figure}[tp]
\centering
\includegraphics[width=0.5\textwidth]{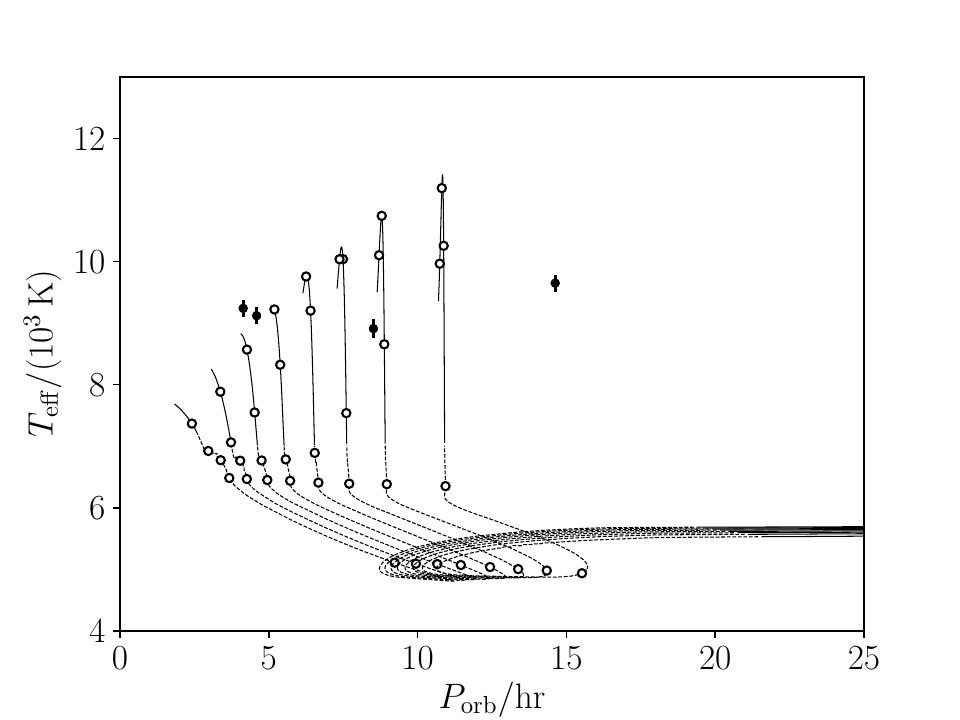}
\caption{The evolutionary tracks of $M_{\rm d,i}=1.0\, M_{\odot}$, $M_{\rm a,i}=0.45\,M_{\odot}$ and $P_{\rm orb,i}$= 2.3, 2.35, 2.4, 2.45, 2.5, 2.55, 2.6, 2.67 days from right to left in the $T_{\rm eff}$ vs. $P_{\rm orb}$ plane. The hollow dots are placed at 1 Gyr intervals. The solid line gives the out of contact of the system, the dashed lines gives the in contact of the system.}
\label{fig:Teff_P_10_045.pdf}
\end{figure}

Figure \ref{fig:Teff_P_10_045.pdf} shows $T_{\rm eff}$ versus $P_{\rm orb}$, and should be compared to the fiducial model in Figure \ref{Teff_P_13_060.pdf}. First note that there are no tracks which end at $P_{\rm orb}>10$ hours due to unstable MT. In the fiducial case, the heavier WDs with shell flashes would end in that region. Given sufficient time, the tracks at $P_{\rm orb}<5$ hours would have slightly larger maximum $T_{\rm eff}$, and would explain the data points better. However, there is insufficient time to reach the elbow.

\subsection{The Maximum ELM WD Progenitor Mass}

Given that more massive donors have a shorter MS phase, this leaves more time for the resultant ELM WDs to cool to $T_{\rm eff} \simeq 9,000\rm K$ and become pulsators. However, sufficiently massive progenitors produce helium cores at terminal age MS which are larger than the maximum ELM WD to avoid shell flashes. Hence there is a maximum progenitor which can create an ELM WD. This section describes models with donor mass $M_{\rm d}=1.5\, M_\odot$ which approaches this limit.

\begin{figure}[tp]
\centering
\includegraphics[width=0.5\textwidth]{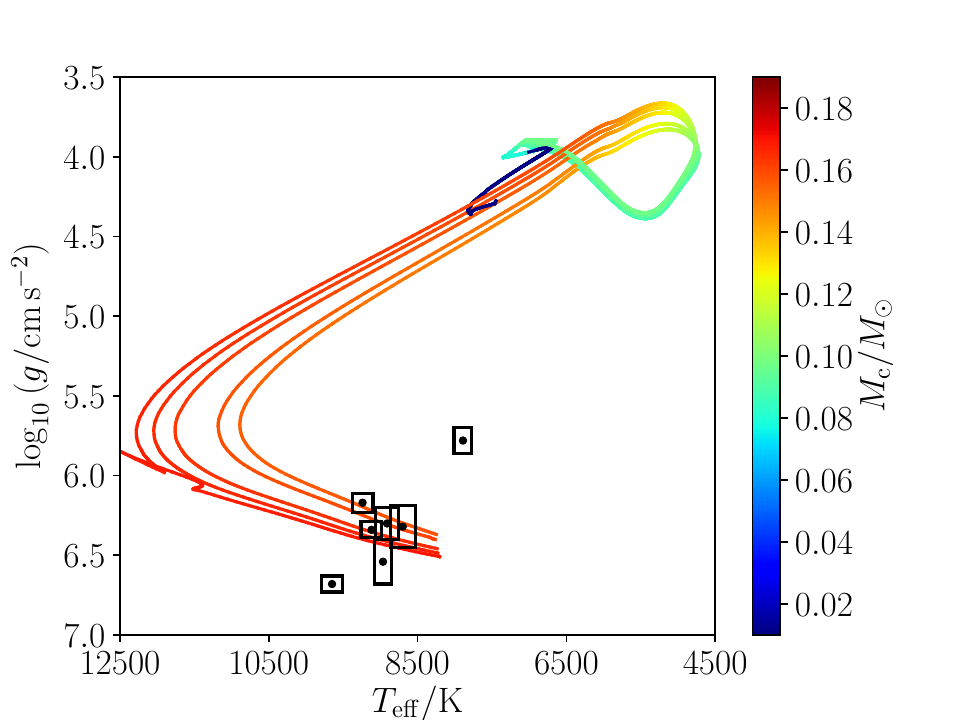}
\caption{Evolutionary models for $M_{\rm d,i}=1.5\, M_{\odot}$ and $M_{\rm a,i}=0.60\, M_{\odot}$. The figure shows the entire range of ELM WDs (all of which have masses greater than 0.16 $M_{\odot}$), covering the range of initial orbital periods $P_{\rm orb,i}=$0.85, 0.86, 0.87, 0.88, 0.89, 0.9 day, from right to left. See Figure \ref{fig:many tracks from RLOF} for the description the black dots. The leftmost track ($P_{\rm orb,i}=$0.9 day) experiences a weak hydrogen flash prior to the WD cooling phase.}
\label{fig:logg_Teff_color_15_060.pdf}
\end{figure}

Figure \ref{fig:logg_Teff_color_15_060.pdf} shows evolutionary tracks for $M_{\rm d,i}=1.5\, M_{\odot}$ and $M_{\rm a,i}=0.6\, M_{\odot}$, for the range of initial orbital periods $P_{\rm orb,i}=$ 0.85, 0.86, 0.87, 0.88, 0.89, 0.9 day. Comparing to the fiducial case in Figure \ref{fig:many tracks from RLOF}, the $M_{\rm d,i}=1.5\, M_{\odot}$ case does not produce lower mass $M_{\rm d,f}$ which cover the small ${\rm log}\,g$ and $T_{\rm eff}$ part of the plot.
In the $M_{\rm d,i}=1.3\, M_\odot$ case, $0.07 \la M_{\rm c}/M_\odot \la 0.1$ before the onset of RLOF for cases which make an ELM WD with $M_{\rm d,f} \la 0.17\, M_\odot$.  By contrast, the runs with $M_{\rm d,i} = 1.5 M_{\odot}$ failed to produce an ELM WD (for which MT ceased) with mass less than 0.168 $M_{\odot}$. The leftmost track in Figure \ref{fig:logg_Teff_color_15_060.pdf} has $P_{\rm orb,i} = 0.85$ days, and initial periods shorter than this value will have continuous MT and never emerge as an ELM pulsator.

\begin{figure}[tp]
\centering
\includegraphics[width=0.5\textwidth]{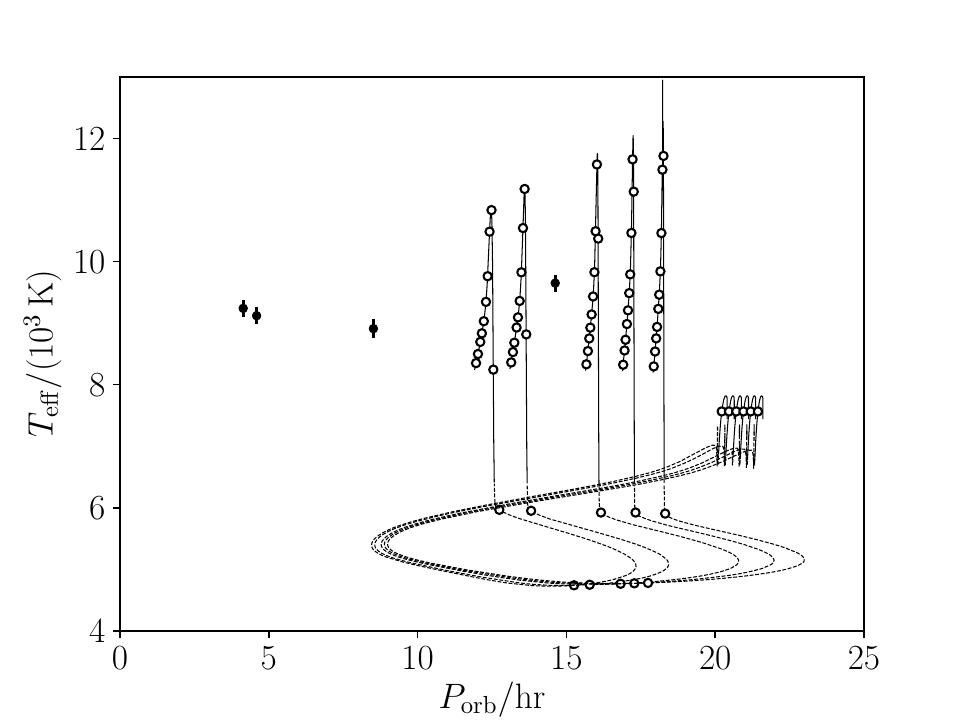}
\caption{Same as Figure \ref{Teff_P_13_060.pdf} for $M_{\rm d,i}=1.5\, M_{\odot}$, $M_{\rm a,i}=0.6\,M_{\odot}$ and $P_{\rm orb,i}$= 0.85, 0.86, 0.87, 0.88, 0.89, 0.9 day from right to left. }
\label{fig:Teff_P_15_060.pdf}
\end{figure}

Figure \ref{fig:Teff_P_15_060.pdf} shows $T_{\rm eff}$ versus $P_{\rm orb}$. The tracks start at the ZAMS with $20 \la P_{\rm orb,i}/{\rm hr} \la 21$, and the MT starts roughly 1-2 Gyr into the MS evolution of the donor star. The final WD thus has more time to cool, more time passes between the start of MT and the end of the 13.7 Gyr simulation, than in the lower-mass donor case. 
After MT commences and enough mass has been lost that $M_{\rm d} \la M_{\rm a}$, the orbit expands dramatically and can exceed the initial separation $a_{\rm i}$. The smallest $P_{\rm orb,f}$ is near 12 hours. As such, the $M_{\rm d,i} = 1.5 M_{\odot}$ case is unable to account for the three systems with $P_{\rm orb} < 12$ hrs.

Further increase of the donor mass above $M_{\rm d,i}=1.5\, M_\odot$ would lead to larger $M_{\rm c}$ at terminal age MS, and larger final WD mass. The upper limit for $M_{\rm d,i}$ which may produce an ELM WD is thus near $1.5 \leqslant M_{\rm d,i}/M_{\odot}\leqslant 1.6$.

\subsection{Mode Periods}
\label{sec:Instability Strip and Seismology}

Adiabatic mode periods have been computed using the GYRE code \citep{2013MNRAS.435.3406T}, which is part of the MESA distribution.

\begin{figure}[tp]
\centering
\includegraphics[width=0.5\textwidth]{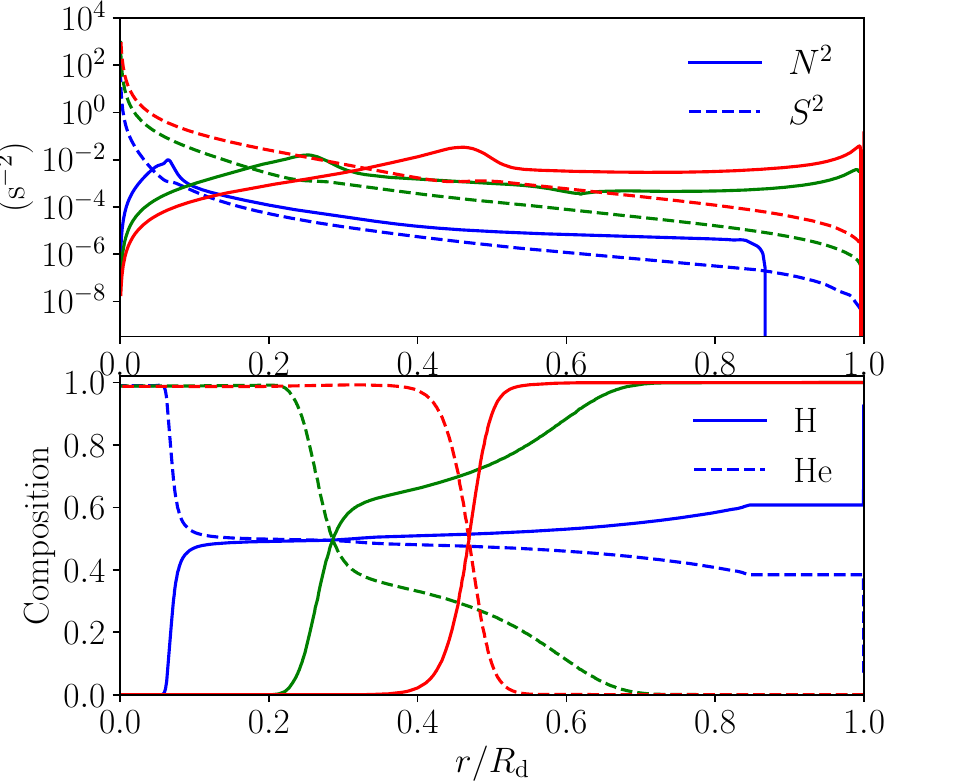}
\caption{Propagation diagram (top panel) for the evolutionary track with $M_{\rm d,i} = 1.3\,M_{\odot}$, $M_{\rm a,i} = 0.6\,M_{\odot}$ and $P_{\rm orb,i}=0.95\,{\rm day}$. The solid lines show the Brunt - V{\"a}is{\"a}l{\"a} frequency while the dashed lines give the square of the Lamb frequency for $\ell=1$. The bump in the buoyancy frequency is due to the composition change from hydrogen to helium with depth. The color of the lines indicate the age, with the blue, green, and red lines representing models at 5.63 Gyr (right after the MT), 7.85 Gyr (at the elbow), and 13.7 Gyr (the termination of the simulation), respectively.}
\label{NS_r.pdf}
\end{figure}

Figure \ref{NS_r.pdf} shows the propagation diagram and the composition versus radius fraction $r/R_{\rm d}$ during the post-MT evolution of the $M_{\rm d,i}=1.3\, M_\odot$, $M_{\rm a,i} = 0.6\,M_{\odot}$ and $P_{\rm orb,i}=0.95\,{\rm day}$ model. Three different times are shown, where the color of the lines indicates the age, with the blue, green, and red lines representing models at 5.63 Gyr (right after the MT), 7.85 Gyr (at the elbow), and 13.7 Gyr (the termination of the simulation), respectively.
The bump in the square of the Brunt - V{\"a}is{\"a}l{\"a} frequency $N^2$ is caused by the composition switch from hydrogen to helium.   
After MT ends, the size of the helium core increases due to sinking of helium in the envelope and ongoing burning of hydrogen in the envelope.

In order for the composition profile to be in diffusive equilibrium, the diffusion timescale must be shorter than the nuclear burning and cooling timescales. The composition profile in Figure \ref{NS_r.pdf} is far from diffusive equilibrium just after MT (blue line) and also at the elbow (green line). The red line is in diffusive equilibrium to a good approximation, and a range of ages (not shown here) were in diffusive equilibrium as well. A close examination of the composition profiles at different ages shows that the residual hydrogen burning ends at nearly the time that diffusive equilibrium is established. The age at which diffusive equilibrium is established was determined for each of the tracks in Figure \ref{fig:logg_Teff_13_060.pdf}, and their position marked by a black cross.  The rightmost track with the shortest $P_{\rm orb,i}$ didn't reach diffusive equilibrium before the second MT phase. The six pulsators with $8,500 \la T_{\rm eff}/{\rm K} \la 9,000$ are in diffusive equilibrium to a good approximation.

\begin{figure}[tp]
\centering
\includegraphics[width=0.5\textwidth]{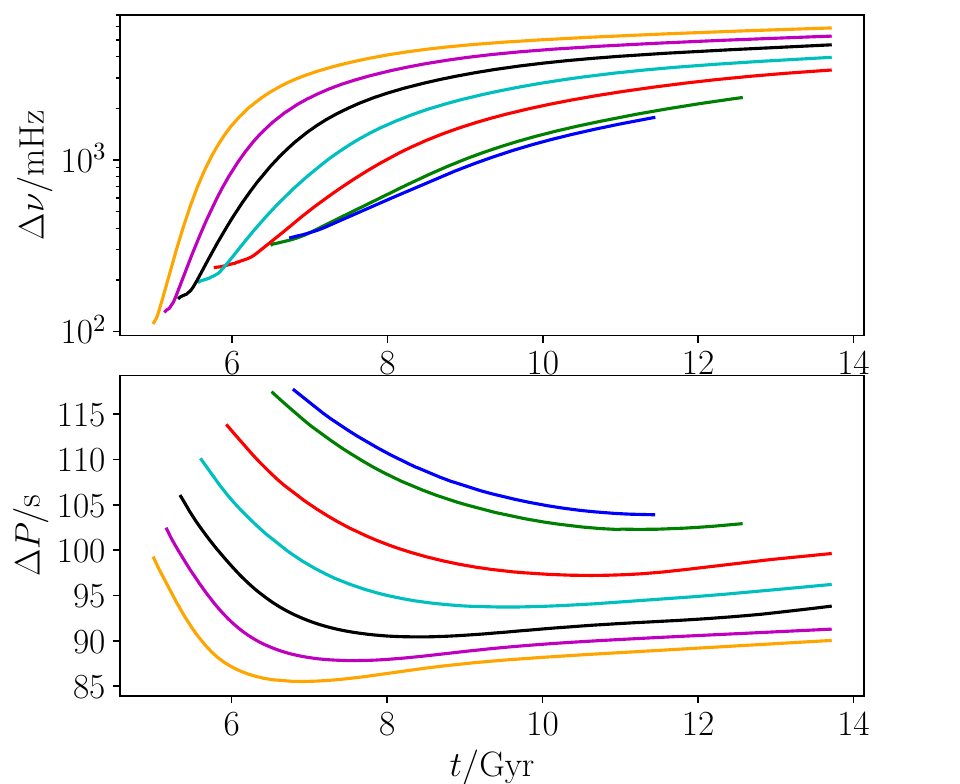}
\caption{The p-mode frequency spacing (top panel) and the g-mode period spacing (bottom panel) versus donor star age. The fiducial model with 
$M_{\rm d,i} = 1.3\,M_{\odot}$ and $M_{\rm a,i} = 0.6\,M_{\odot}$ is used, with $P_{\rm orb,i}=$ 0.90 (blue), 0.91 (green), 0.93 (red), 0.95 (cyan), 0.97 (black), 0.99 (magenta) and 1.01 (orange) days.}
\label{pgspc_age.pdf}
\end{figure}

Fig.~\ref{pgspc_age.pdf} shows the p-mode frequency spacing and g-mode period spacing for the fiducial model after MT has ceased.
The different color lines represent different $P_{\rm orb,i}$. The blue and green lines appear shorter because they were terminated at the start of a second phase of MT. The g-mode period spacing is strongly dependent on the WD mass, so the lines differ by up to 30\%. Also, g-mode period spacing depends on the age, mainly through the thickness of the hydrogen envelope, so there can be about $15\%$ differences in the period spacing along an individual evolutionary track. A minimum in the g-mode period spacing occurs near the elbow separating the pre-WD and the WD cooling track. The p-mode frequency spacing becomes nearly constant for the high-mass models, however for the lower mass models the spacing is slowly increasing in time over many Gyrs.

\begin{figure}[tp]
\centering
\includegraphics[width=0.5\textwidth]{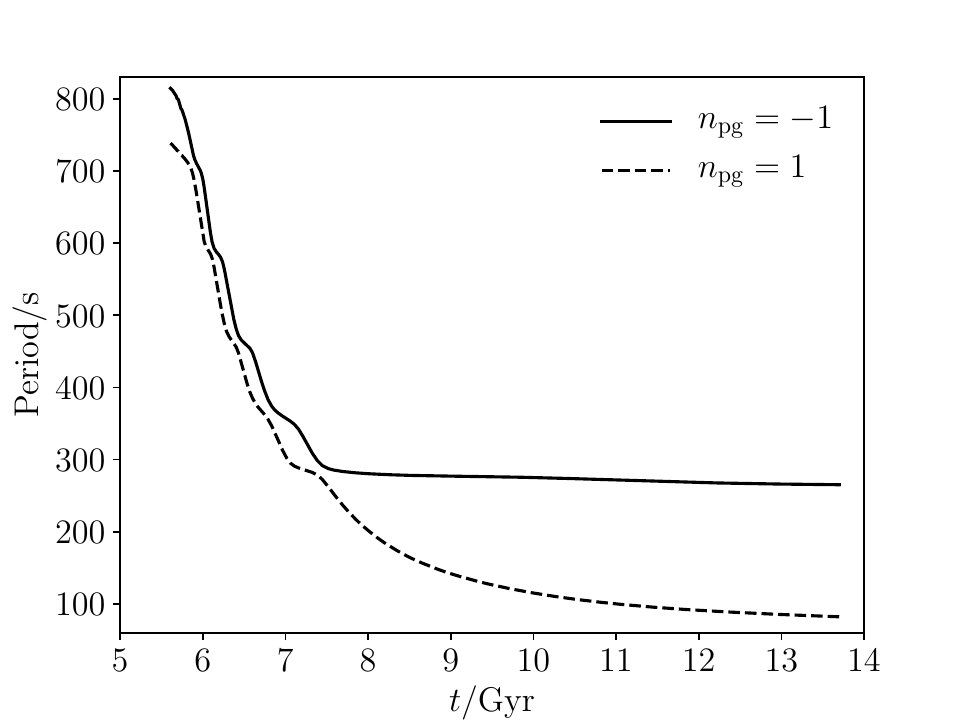}
\caption{The lowest order of p-mode (dashed line) and g-mode (solid line) periods versus age after the MT phase for the track with $M_{\rm d,i}=1.3M_{\odot}$, $M_{\rm a,i}=0.6M_{\odot}$, $P_{\rm orb,i}=0.95$ day.}
\label{pgmode_age.pdf}
\end{figure}

Fig.~\ref{pgmode_age.pdf} displays the lowest order g-mode and p-mode for one evolutionary track ($M_{\rm d,i} = 1.3\,M_{\odot}$, $M_{\rm a,i} = 0.6\,M_{\odot}$, $P_{\rm orb,i}= 0.95$ day). Most of the oscillation modes are mixed modes on the pre-WD track, meaning that near the radiative core, the mode behaves like a g-mode, and in the outer convection zone the mode behaves like a p-mode with larger radial displacement. A sequence of avoided crossings are observed during the approximately 2 Gyr pre-WD phase. Starting at 7.5 Gyr, on the WD cooling track, the avoided crossings end, and the g-mode and p-mode are distinct, separated with gap in period. This period separation increases during the subsequent WD cooling phase.

\begin{figure}[tp]
\centering
\includegraphics[width=0.5\textwidth]{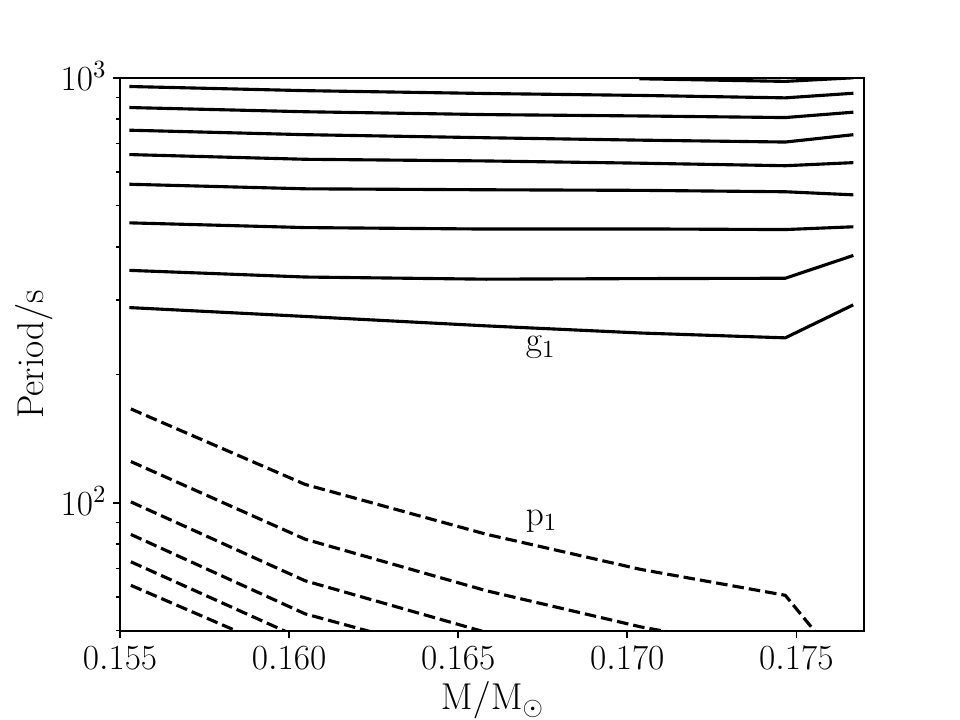}
\caption{Eigen-periods versus the WD mass for different models evaluated at $T_{\rm eff} = 9000$ K. The six models with $M_{\rm d,i} = 1.3\,M_{\odot}$ and $M_{\rm a,i} = 0.6\,M_{\odot}$, and $P_{\rm orb,i}=$ 0.93, 0.95, 0.97, 0.99, 1.01, 1.03 days  are used, and the mode periods evaluated for the model with $T_{\rm eff}$ closest to 9000K.}
\label{freq_mass.pdf}
\end{figure}

Fig.~\ref{freq_mass.pdf} shows the $\ell=1$ p-modes (dashed) and g-modes (solid) at a fixed $T_{\rm eff} = 9000$ K. The final WD masses are from six evolutionary tracks, with $P_{\rm orb,i}=$ 0.93, 0.95, 0.97, 0.99, 1.01, 1.03 days and $M_{\rm d,i} = 1.3\,M_{\odot}$, $M_{\rm a,i} = 0.6\,M_{\odot}$. The period gap between the p-modes and g-modes increases with the WD mass. The g-mode periods decrease slightly with the increased final mass in the mass range of the ELM WD. For WDs with masses above about $0.18 M_{\odot}$, the gap between the p-modes and g-modes begins to increase even more rapidly. The mode periods decrease with increasing WD mass, which agrees with CA's result \citep{2014A&A...569A.106C}.


\section{ Discussion }
\label{sec:discussion}

\subsection{Pre-WD Structure and Orbital Periods just after Mass Transfer Ends}

\begin{figure}[tp]
\centering
\includegraphics[width=0.5\textwidth]{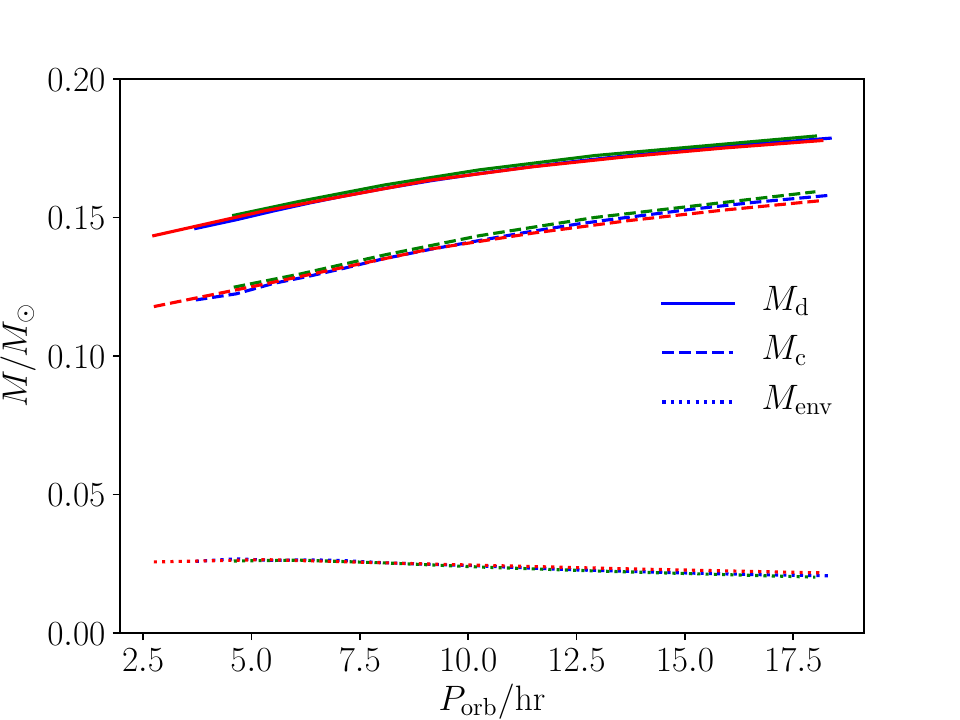}
\caption{Total mass ($M_{\rm d}$), helium core mass ($M_{\rm c}$) and envelope mass ($M_{\rm env}=M_{\rm d}-M_{\rm c}$) as a function of $P_{\rm orb}$ at the end of the first phase of MT. The results of $M_{\rm d,i}=1.3\,M_{\odot}$, $M_{\rm a,i}=0.6\,M_{\odot}$ and $0.9 \leqslant P_{\rm orb,i}/{\rm day} \leqslant 1.03$ with a step of 0.01 day is in blue. The results from $M_{\rm d,i}=1.3\,M_{\odot}$, $M_{\rm a,i}=0.9\,M_{\odot}$ and $0.78 \leqslant P_{\rm orb,i}/{\rm day} \leqslant 0.89$ is in green. The results of $M_{\rm d,i}=1.1\,M_{\odot}$, $M_{\rm a,i}=0.6\,M_{\odot}$ and $0.84 \leqslant P_{\rm orb,i}/{\rm day} \leqslant 1.40$ is in red. See Figures \ref{fig:many tracks from RLOF}, \ref{tracks13_060.pdf}, \ref{Masses13_Porb.pdf} and \ref{q_vs_mass.pdf}. }
\label{masses_Porb.pdf}
\end{figure}

Figure~\ref{masses_Porb.pdf} shows the stellar mass as a function of $P_{\rm orb}$ just after the MT phase has ended. Three different initial binary mass configurations (fiducial case, larger accretor mass, and smaller donor mass) are plotted. The helium core mass and hydrogen envelope mass are also plotted for each of these systems. Keep in mind that $P_{\rm orb}$ continues to change in the pre-WD and the WD cooling phases due to gravitational wave losses.Also, there is continued burning of the envelope adding to the core.

For the fiducial case, $P_{\rm orb,i}$ smaller than 0.9 day results in continuous accretion and thus no ELM WD pulsator. 
The lowest ELM WD mass is $M_{\rm d,f}=0.146\, M_{\odot}$, with $M_{\rm c}=0.120\, M_{\odot}$ and $P_{\rm orb}=$ 3.72 hours. The envelope is 18\% of the total mass in this case, much larger than for a standard $0.6\, M_\odot$ carbon/oxygen WD. A large fraction of the radius of the star is also taken up by the envelope in this case. 
All else being equal, larger $P_{\rm orb,i}$ results in higher pre-WD masses and $M_{\rm c}$, but lower $M_{\rm env}$, immediately post-MT. For $P_{\rm orb,i}$ above 1.03 days (in the fiducial case), the WD experiences hydrogren flashes prior to the cooling phase. At this upper boundary, the pre-WD mass is 0.179 $M_{\odot}$ and $P_{\rm orb}$ is 18.35 hours immediately post-MT, with an envelope containing only 11.5\% of the total mass, which is smaller than all the ELM WDs with no hydrogen flashes before cooling. For non-ELM WDs which experience hydrogen flashes prior to the cooling phase, the resulting envelope is much thinner than for the lower-mass ELM WD. The trend of the thinner envelope with an increasing total WD mass agrees with \cite{2016A&A...595A..35I}.

The results from a simulation with high-mass companion, $M_{\rm a,i}=0.9\, M_{\odot}$, are plotted as green lines in Figure \ref{masses_Porb.pdf}. The blue and green lines are nearly overlapping, producing ELM WDs with almost identical mass, composition, and orbits immediately post-MT. Similarly, the results for a lower-mass donor $M_{\rm d,i} = 1.1 M_{\odot}$ are shown in red; this setup can create ELM WDs with even lower masses and shorter $P_{\rm orb}$. The minimum ELM WD mass for this setup is 0.143 $M_{\odot}$, with $P_{\rm orb}=$ 2.75 hours immediately post-MT, and the range of $P_{\rm orb,i}$ that results in ELM WDs is considerably larger than for higher-mass donors. For small donor masses (e.g., $M_{\rm d,i} = 1.0$ and 1.1 $M_{\odot}$), the thick convective envelope present during the MS phase causes magnetic braking to be much stronger, resulting in a wider accessible range of $P_{\rm orb,i}$. \cite{2017MNRAS.467.1874C} used a wider range for $M_{\rm d}$, $M_{\rm a}$ and $P_{\rm orb}$ with different metallicities. Their mass-period relation is in agreement with Figure~\ref{masses_Porb.pdf}. And the difference in metallicity doesn't affect this relation at the low-mass WD range.

\subsection{Models Producing Higher $T_{\rm eff}$ at Shorter $P_{\rm orb}$}

From Figure \ref{Teff_P_13_060.pdf}, two of the ELM WDs, J1840 and J1112, have $4 \la P_{\rm orb}/{\rm hr} \la 5$ (a range that is accessible with our simulations), but with a higher $T_{\rm eff}$ that falls slightly above the theoretical tracks. This section is about making ELM WDs with $T_{\rm eff} \approx$ 9000 K and short orbital periods $P_{\rm orb} \la 5$ hr.

\begin{figure}[tp]
\centering
\includegraphics[width=0.5\textwidth]{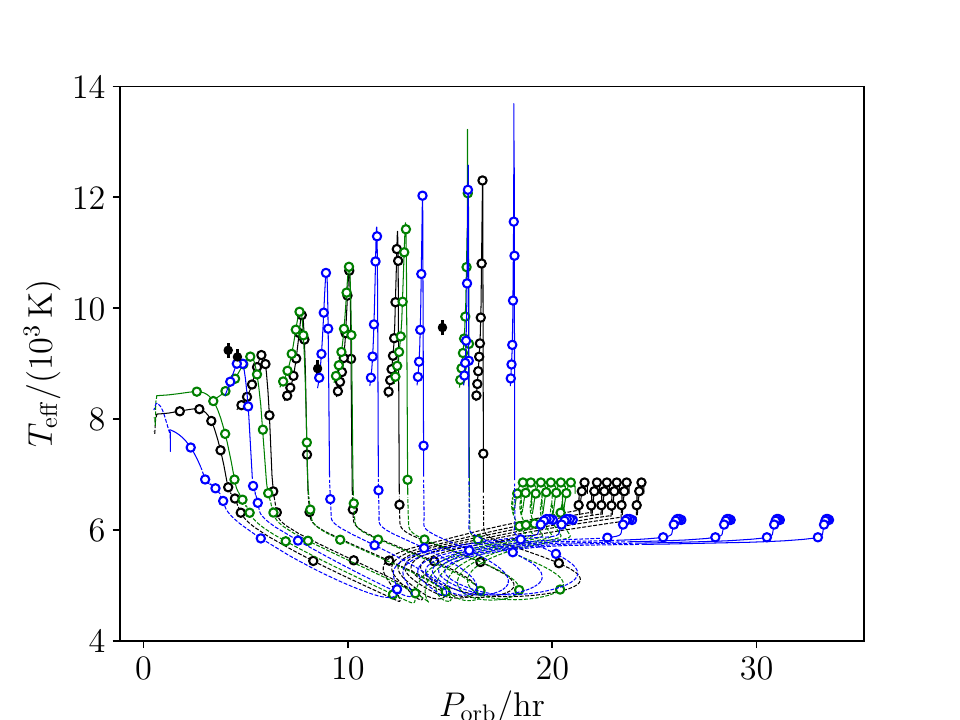}
\caption{ Evolutionary tracks for trying different donor and accretor mass with $M_{\rm d,i}=1.1\,M_{\odot}$, $M_{\rm a,i}=0.6\,M_{\odot}$ and $P_{\rm orb,i}=$ 0.84, 0.88, 1.0, 1.1, 1.2, 1.3, 1.4 days in blue; $M_{\rm d,i}=1.3\,M_{\odot}$, $M_{\rm a,i}=0.6\,M_{\odot}$ and $P_{\rm orb,i}=$ 0.90, 0.93, 0.95, 0.97, 0.99, 1,02 days in black; and $M_{\rm d,i}=1.3\,M_{\odot}$, $M_{\rm a,i}=0.9\,M_{\odot}$ and $P_{\rm orb,i}=$ 0.78, 0.80, 0.82, 0.84, 0.86, 0.88 day in green. One of the blue tracks ($M_{\rm d,1}=1.1\,M_{\odot}$, $M_{\rm a,i}=0.6\,M_{\odot}$, $P_{\rm orb,i}=0.88\,{\rm day}$) passes though the observed $T_{\rm eff}$ and ${\rm log}\,g$ from observation.}

\label{fig:Teff_P_compare.pdf}
\end{figure}

Figure~\ref{fig:Teff_P_compare.pdf} compares evolutionary tracks for simulations with different $M_{\rm d,i}$ and $M_{\rm a,i}$. Prior to the WD cooling phase, the donor star reaches a maximum $T_{\rm eff}$.  The trend is for this maximum $T_{\rm eff}$ to decrease with decreasing $P_{\rm orb,i}$, and for the fiducial case it appears that having $T_{\rm eff} \approx$ 9000 K with $4 \la P_{\rm orb}/{\rm hr} \la 5$ is inaccessible. For a higher accretor mass, and even more dramatically for a lower donor mass, the maximum $T_{\rm eff}$ is increased relative to the fiducial case. So to make a WD with $T_{\rm eff} >$ 9000 K at short orbital periods, the preference is to have a low-mass donor and high mass accretor.

\subsection{Stellar Engineering Construction of ELM WD}

Instead of making ELM WDs models through binary evolution including magnetic braking, a simpler and cheaper alternative would be the following. Evolve a single ZAMS star until it reaches the desired $M_{\rm c}$. Then rapidly (on a timescale much shorter than the thermal and nuclear burning timescales) remove the envelope until the desired $M_{\rm env}$ is left. The resulting model would represent the start of the pre-ELM WD track seen in this work. The two parameters are $M_{\rm c}$ and $M_{\rm env}$, for a fixed composition. The star is then evolved through the pre-ELM WD and WD cooling tracks.

Though much simpler, the problem with this method is that it is not known a priori what to choose for $M_{\rm c}$ and $M_{\rm env}$. Furthermore, this method does not give the expected $P_{\rm orb}$ for the binary, or possible ranges of the mass of the companion WD. The former issue has been addressed in Figure \ref{masses_Porb.pdf}, which shows how the total stellar mass is partitioned into core and envelope. This greatly restricts the range of allowed $M_{\rm env}$ because even far larger $M_{\rm env}$, up to half the mass of the star can be used for $M<0.17\, M_\odot$ without incurring shell flashes. Such large $M_{\rm env}$ would have many Gyrs of hydrogen burning until a more physically-motivated envelope size would result.

\section{ Conclusions }
\label{sec:conclusions}

This work has discussed the formation of double WD binaries in which one of the stars is an ELM WD with mass $M\la 0.18\, M_\odot$. The main results of the paper are as follows.

-- ELM WDs cannot be formed via conservative MT. The mass-loss rate for conservative MT is not fast enough to remove more than 90\% of the WD progenitor star before the helium core has grown beyond the ELM WD regime. As a result, the minimum mass of an ELM WD made by conservative MT is about 0.2 $M_{\odot}$.

--ELM WDs are not likely to be formed through common envelope evolution.  For donor and accretor masses consistent with producing an ELM WD, the  binding energy of the donor's envelope (90\% of the donor mass) is so large that the final binary orbital separation would be unphysically small (i.e., smaller than the stellar radii), implying merging of the two stars.

-- The ELM WD binary formation pathway investigated in this paper posits that the ELM WD progenitor is the initially less massive star.  In this picture, the first MT phase occurs when the more massive companion evolves off the MS and the binary enters a common envelope phase. Upon ejection of the common envelope material, the initially more massive star becomes a helium or a carbon-oxygen WD. Once the initially less massive star evolves off the MS, a second (RLOF) MT phase takes place.  The donor star in this phase is the progenitor of the ELM WD. Magnetic braking during this phase is crucial to strip the envelope before the core grows too large. Subsequent gravitational wave angular momentum losses lead to decay of the shortest orbital period systems, which may have a second phase of MT. MT was assumed to be completely non-conservative.

-- The possible mass range for the ELM WD progenitor is $1.0 \leqslant M_{\rm d,i}/M_{\odot}\leqslant 1.5$. For initial stellar mass below 1.0 $M_{\odot}$, the WD cannot reach the WD cooling phase within a Hubble time with the initial Z=0.01. For initial masses greater than 1.5 $M_{\odot}$, the (convective) helium core grows too large to make ELM WDs with mass less than about 0.17 $M_{\odot}$.

-- Similar final ELM WDs can be produced via different combinations of donor mass, accretor mass and initial orbital period. In the first case the increasing donor radius as it evolves off the MS triggers the RLOF, while in the second case the decreasing orbital radius (caused by magnetic braking) shrinks the Roche lobe radius to the point where it reaches the stellar radius of the donor. An ELM WD binary with short $P_{\rm orb}$ and high $T_{\rm eff}$ may be produced from a low-mass donor with high-mass accretor. In general, the accretor mass should be large enough to avoid unstable MT.

-- The mass range of ELM WDs created via RLOF is $0.146 \la M/M_\odot \la 0.18$, with $2 \leqslant P_{\rm orb}/{\rm hr} \leqslant 20$. For higher mass WDs there can be several hydrogen flashes prior to the cooling phase, and the final $P_{\rm orb}$ is wider than for ELM WDs with no shell flashes.

\acknowledgements

We thank Lars Bildsten, Bill Wolf and Bill Paxton for useful discussion on stable and unstable MT to make ELMs. We also thank Morgan Macleod for ideas about the common envelope phase. We thank Christian Hayes for discussing the metallicity in disk and halo stars. Meng Sun thanks Dom Pesce for the suggestions and carefully revising the paper. We thank Christopher Tout referee for comments and suggestions which have improved this paper. This work was supported by NASA ATP grant NNX14AB40G.
 

\appendix
Appendix \ref{sec:conservative} and \ref{sec:ce} discuss WD formation by conservative MT and CE evolution. It is shown that neither of these channels are likely to form an ELM WD.
\bigskip

\section{Appendix A Conservative Evolution}
\label{sec:conservative}

The simplest case to consider for binary interaction is conservative MT with constant total mass and orbital angular momentum. Evolution occurs on the nuclear timescale of the donor star, and significant orbit expansion occurs as it ascends the RGB.

\citet{2000MNRAS.319..215H} extended earlier studies (e.g. \citealt{1967ZA.....66...58K}) by considering $Z=0.02$ stars with a range of ZAMS donor mass $1 \leqslant M_{\rm d,i}/M_{\odot} \leqslant 8$ and mass ratios $1.1 \leqslant (q=M_{\rm d,i}/M_{\rm a,i}) \leqslant 4$. Here $M_{\rm d,i}$ and $M_{\rm a,i}$ are the initial donor and accretor masses. The initial orbital separation and period, $a_{\rm i}$ and $P_{\rm orb,i}$, were set so that RLOF commenced in the early, middle or late Hertzsprung gap. This study found that the smallest WD masses are produced through a combination of the smallest possible donor masses, which evolve in the age of the Galaxy, the smallest accretor masses, to give higher mass-loss rates, and the smallest initial separations, to avoid building up the helium core. For a given donor mass, there is a limit on how small the accretor mass can be in order to avoid unstable MT. The smallest mass WD in their 150 simulations was $M=0.21\, M_\odot$, with parameters $M_{\rm d,i}=1.0\, M_\odot$, $M_{\rm a,i}=0.5\, M_\odot$ and $P_{\rm orb,i}=0.49\, {\rm day}$. Hence while conservative evolution produces masses approaching the ELM WD mass range, it appears that it cannot robustly produce WDs in the mass range $0.1\leqslant M/{M_\odot} \leqslant 0.2$. Further, the final orbital periods have $P_{\rm orb,f} \approx 1\, {\rm week}$, much wider than the observed ELM systems.

\begin{figure}[tp]
\centering
\includegraphics[width=0.5\textwidth]{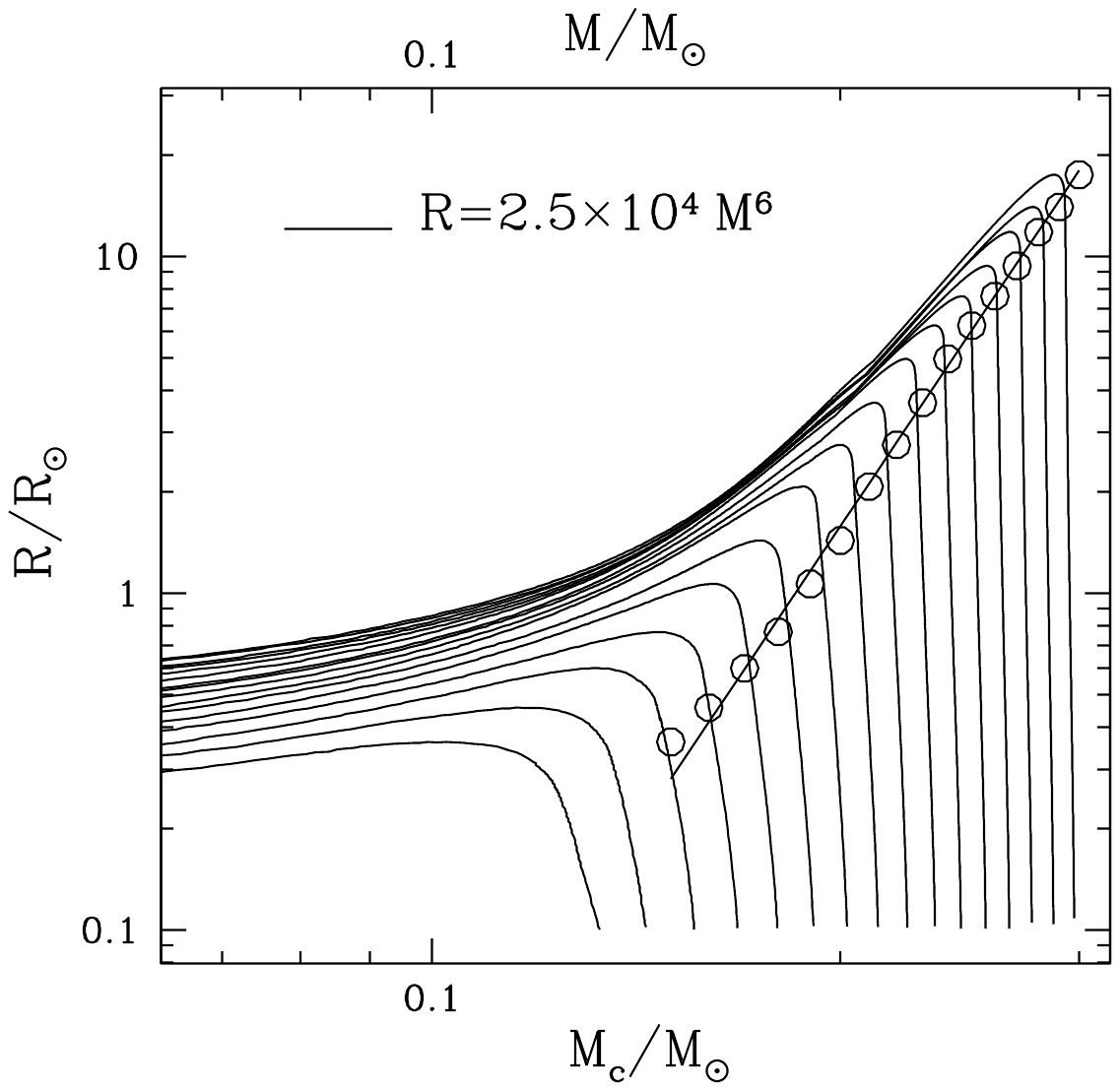}
\caption{ Solid lines show stellar radius ($R$, abscissa) versus helium core mass ($M_{\rm c}$, lower axis ordinate) during the evolution of single stars with metallicity $Z=0.01$. The lines represent stellar masses $M/{M_\odot}=0.15,0.16,...,0.30$ from bottom to top. The open circles show the maximum radius along each track versus total mass ($M$, upper axis ordinate). The solid line is a fit to the circles, given by $R/R_\odot=2.5\times 10^4\, (M/M_\odot)^6$.}
\label{fig:R_vs_M}
\end{figure}

The numerical results of \citet{2000MNRAS.319..215H} can be understood with the analytic treatment in \citet{1971A&A....13..367R}. For conservative evolution, the total mass $M_{\rm d,f} + M_{\rm a,f} = M_{\rm d,i} + M_{\rm a,i}$ is constant between the initial and final states, where $M_{\rm d,f}$ and $M_{\rm a,f}$ are the final donor and accretor star masses. The constancy of orbital angular momentum implies that $a_{\rm i} (M_{\rm d,i}M_{\rm a,i})^2=a_{\rm f} (M_{\rm d,f}M_{\rm a,f})^2$, where $a_{\rm f}$ is the final separation after the MT. Kepler's 3rd law can be used to write $a_{\rm i} = (G(M_{\rm d,i} + M_{\rm a,i})P_{\rm orb,i}^2/(2\pi)^2)^{1/3}$, where $P_{\rm orb,i}$ is the orbital period where RLOF commences. Lastly, a mass--radius relation is required for the low-mass RGB star, at the maximum radius attained before the envelope becomes too thin and the radius shrinks. Following \citet{1971A&A....13..367R}, this is estimated from single star evolution tracks.

MESA was used to evolve stars of constant total mass $M/M_\odot = 0.15,0.16,...,0.30$ from ZAMS to the first shell flash on the WD cooling track, as shown in Figure \ref{fig:R_vs_M}. A maximum occurs in radius, beyond which the radius shrinks with further decrease of the envelope. The open circles show maximum radius versus total mass, and the solid line is a fit given by $R(M_{\rm d,f}) = 2.5\times 10^4\, R_\odot (M_{\rm d,f}/M_\odot)^6$. Combining all these results, and approximating $M_{\rm a,f} = M_{\rm d,i}+M_{\rm a,i}-M_{\rm d,f} \simeq M_{\rm d,i}+M_{\rm a,i}$, gives the final WD mass 

\be
M_{\rm d,f} & = & 0.29\, M_\odot \left( \frac{P_{\rm orb,i}}{1\, \rm day} \right)^{0.087} \left( \frac{M_{\rm d,i}M_{\rm a,i}}{M_\odot (M_{\rm d,i}+M_{\rm a,i})} \right)^{0.26}.
\label{eq:mwd_conservative}
\ee
\citet{2000MNRAS.319..215H}'s conclusions about the variation of $M_{\rm d,f}$ with $P_{\rm orb,i}$, $M_{\rm d,i}$ and $M_{\rm a,i}$ are directly observed in this formula. It agrees with the final WD masses of \citet{2000MNRAS.319..215H} to an accuracy of 2-3\%.

To derive the smallest possible WD mass from conservative evolution, the accretor mass is evaluated at the stability limit $M_{\rm a,i} \simeq M_{\rm d,i}/2.5$, the donor mass is set to the smallest that can evolve in the age of the Galaxy, $M_{\rm d,i} \simeq 1.0\, M_\odot$ for $Z=0.01$, and the initial orbital period is set so that RLOF commences near the end of the MS, $P_{\rm orb,i} \simeq 0.6\, \rm day$, with the result

\be
M_{\rm d,f,min} & \simeq & 0.20\, M_\odot.
\ee
Hence conservative evolution cannot lead to an ELM WD of mass $M \la 0.18\, M_\odot$.

\begin{figure}[tp]
\centering
\includegraphics[width=0.5\textwidth]{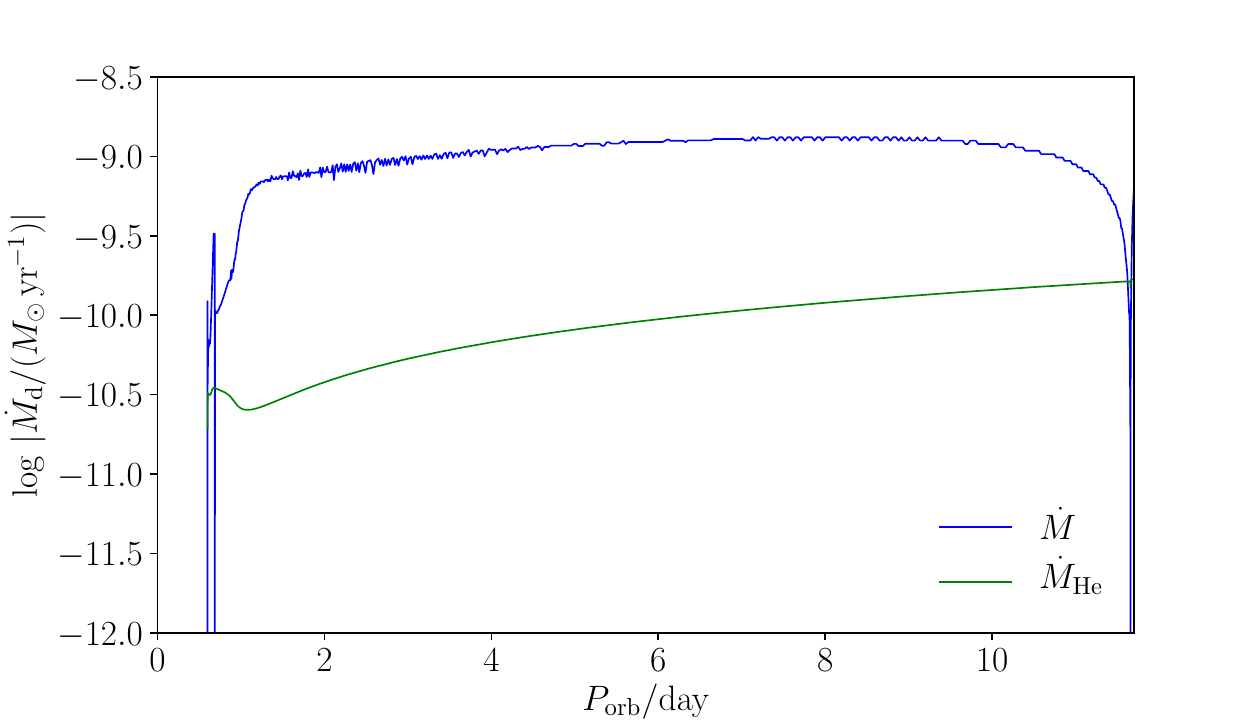}
\caption{Mass loss rate, $\dot{M}_{\rm d}$, and growth rate of the helium core, $\dot{M}_{\rm c}$, for conservative evolution of a $M_{\rm d,i}=1.2\, M_\odot$ donor with a $M_{\rm a,i}=0.8\, M_\odot$ accretor, and initial orbital period $P_{\rm orb,i}=0.6\, {\rm day}$. The solid line is the mass-loss rate from the star, and the dashed line is the growth rate of the helium core. 
}
\label{Mdot_P.pdf}
\end{figure}

It is instructive to consider why conservative MT produces WDs with mass $M>0.2\, M_\odot$. Consider a donor of mass $M_{\rm d,i} = 1.2\, M_\odot$ near the end of the MS, with a $M_{\rm c} \simeq 0.1\, M_\odot$ core already built. For the core to be limited to $M_{\rm c} \la 0.15\, M_\odot$ means that only $0.05\, M_\odot$ can be added to the core while $1.05\, M_\odot$ must be lost by RLOF, requiring a mass-loss rate for donor of $|\dot{M}_{\rm d}| \ga 20\, \dot{M}_{\rm c}$, where $\dot{M}_{\rm c}$ is the rate at which the helium core grows due to hydrogen shell burning. Figure \ref{Mdot_P.pdf} shows a MESA calculation of conservative binary evolution with a $M_{\rm d,i}=1.2\, M_\odot$ donor transferring mass to a $M_{\rm a,i}=0.8\, M_\odot$ accretor, with initial orbital period of $P_{\rm orb,i}=0.6\, \rm day$. For $M_{\rm d,i}>M_{\rm a,i}$, there is an initial phase of TTMT at high $\dot{M}_{\rm d}$. Once the donor mass nearly equals the accretor mass $M_{\rm d} \la M_{\rm a}$, this is followed by a second phase at lower $\dot{M}_{\rm d}$ on the nuclear timescale of the donor. It is during the second phase that the helium core builds up to large mass. Figure \ref{Mdot_P.pdf} shows that, as shell burning causes the radius to expand, setting $\dot{M}_{\rm d}$, it is also adding to the helium core at a rate $\dot{M}_{\rm c}$.  The second phase has $10 \la |\dot{M}_{\rm d}|/\dot{M}_{\rm c} \la 15$, which allows the core to grow too much. What is needed is a faster rate of RLOF, to limit the increase of $M_{\rm c}$. 

Lastly, conservative evolution tends to produce orbital periods far larger than that of ELM WDs. Plugging the result in Equation \ref{eq:mwd_conservative} in to Kepler's third law, the final orbital period is

\be
P_{\rm orb,f} = &2.2\, {\rm day} \left( \frac{P_{\rm orb,i}}{0.5\, \rm day} \right)^{0.74} \left( \frac{M_{\rm d,i}}{1\, M_\odot}\frac{0.5\, M_\odot}{M_{\rm a,i}} \frac{1.5\, M_\odot}{M_{\rm d,i}+M_{\rm a,i}} \right)^{2.22},
\ee

far larger than that observed for the ELM WDs.

Noise is apparent in the MT rate in Figure \ref{Mdot_P.pdf}. To assess the size of the noise for different values of MESA solver parameters, runs were carried out with smaller values of ``varcontrol\_target” in the MESA namelist. This parameter controls the relative variation in values of the solution from one model to the next. A decrease of varcontrol\_target from $10^{-3}$ to $10^{-4}$ contained smaller amounts of noise by a factor 10 in $\dot{M}_d$, during the time intervals where the MT rate was noisy (i.e., at the start of MT). And this change gave values of $M_c$, $M_{\rm env}$, $R_d$, $T_{\rm eff}$ and $\log_{10}\,g$ to better than  1\%.


\section{ Appendix B common envelope evolution }
\label{sec:ce}

If the progenitor of the ELM WD is too massive compared to the companion, then MT can be unstable and grow to extremely large mass. At such high mass-loss rates, the mass is unable to settle on the accretor, and the donor's ejected envelope forms a common envelope around the uncovered core of the donor and the accretor \citep{2006csxs.book..623T}. Drag forces from the two stars then inject energy and angular momentum into the envelope. If there is sufficient orbital energy to eject the envelope, then the two stars emerge as a much more compact binary. If there is insufficient energy to eject the envelope, merging results.

The problem with forming an ELM WD by CE is that the envelope is much more massive than the core, and an extreme spiral-in is required to eject the envelope. Merging may be the outcome in many cases. Consider a numerical example with a donor star of mass $M_{\rm d}=1.4\, M_\odot$ with He core $M_{\rm c}=0.15\, M_\odot$. For metallicity $Z=0.01$, the donor's radius is $R_{\rm d}=2.6\, R_\odot$ for this core size. For unstable MT, there is an upper limit on the accretor's mass of $M_{\rm a} \la M_{\rm d}/2.5 =0.56\, M_\odot$ for the chosen donor mass. The factor 2.5 was found using MESA simulations for conservative MT, using a range of donor masses. The energy equation for CE evolution equates the binding energy of the donor's envelope to the change in orbital energy \citep{1984ApJ...277..355W}:

\be
\frac{GM_{\rm d}(M_{\rm d}-M_{\rm c})}{\lambda R_{\rm d}} & = & \alpha \left( \frac{GM_{\rm c} M_{\rm a}}{2a_{\rm f}} - \frac{GM_{\rm d} M_{\rm a}}{2a_{\rm i}} \right),
\label{eq:ce_energy}
\ee
where $a_{\rm f}$ is the separation of the resultant binary after CE. At contact, $R_{\rm d}=r_{\rm L}(M_{\rm d}/M_{\rm a}) a_{\rm i}$, where $r_{\rm L}(2.5) \simeq 0.46$ relates the stellar radius to the initial separation for a star in Roche-lobe contact \citep{1983ApJ...268..368E}. Solving for the final separation and plugging in numbers gives

\be
a_{\rm f} & = & R_{\rm d}\ \left( \frac{M_{\rm c} M_{\rm a}}{ r_{\rm L} M_{\rm d} M_{\rm a} + (2/\alpha\lambda) M_{\rm d} (M_{\rm d}-M_{\rm c})} \right)
\nonumber \\ & \simeq & 
 R_{\rm d}\ \left( \frac{\alpha \lambda}{2} \right) \left( \frac{M_{\rm a}}{M_{\rm d}} \right) \left( \frac{M_{\rm c}}{M_{\rm d}-M_{\rm c}} \right)
 \simeq 0.06\, R_\odot.
\ee
The small separation is due to two requirements. First $M_{\rm d}/M_{\rm a} \ga 2.5$ in order to have unstable MT, and secondly the core is much less massive than the envelope so that $M_{\rm c}/(M_{\rm d}-M_{\rm c}) \approx 0.1$.

If the progenitor of the ELM WD was the initially more massive star, then the companion is a MS star of radius $R_{\rm a} \approx 0.5\, R_\odot$, which cannot fit inside the orbit, implying merging. If the progenitor of the ELM was the initially less massive star, and the initially more massive star became a massive WD, after the ELM WD formed, the radius of the massive WD is smaller than the ELM WD. The radius of a $M_{\rm c}=0.15\, M_\odot$ ELM WD with a thick hydrogen envelope can be as large as $0.05 \leqslant R_{\rm ELM}/R_{\odot} \leqslant 0.15$, and so the ELM could not fit inside the final separation, and a merger would result. Back to the numerical example at the begining of this section, for the mass ratio $M_{\rm c}/M_{\rm a}=0.15/0.56=0.27$ and an ELM WD radius of $R_{\rm ELM}=0.08\, R_\odot$, the orbit would have to be wider than $a_{\rm f} = R_{\rm ELM}/r_{\rm L}(0.27) > 0.29\, R_\odot$ for the ELM WD not to be in contact. For the ELM phase to be long-lived against orbital decay by gravitational radiation, the orbital period should be significantly wider. Similarly, the binary population syntheses work by \cite{2009ARep...53..214B} indicates there is less probablity that the low-mass helium WD is formed after the CE phase. \cite{2017MNRAS.467.1874C} show that EL CVn, which is close to the systems in this paper, cannot be produced by CE for the same reason. The orbital separation shrinks too much that the two stars may merge.



\end{document}